\documentclass[aps,prb,twocolumn,superscriptaddress,floatfix,amsmath,showpacs]{revtex4}
\usepackage{ae}
\usepackage{graphicx}
\usepackage{bm}
\usepackage{wasysym}
\usepackage{epsf}
\usepackage{epsfig}
\usepackage{amssymb}

\begin{document}

\title[Elastic line]
{Growing correlations and aging of an elastic line in a random potential}
\vskip 10pt
\author{Jos\'e Luis Iguain}
\affiliation{Instituto de Investigaciones F\'{\i}sicas de Mar del Plata (IFIMAR) and
Departamento de F\'{\i}sica FCEyN,\\
Universidad Nacional de Mar del Plata, De\'an Funes 3350, 7600 Mar del
Plata, Argentina}
\author{Sebastian Bustingorry}
\affiliation{CONICET, Centro At\'omico Bariloche-Comisi\'on Nacional de Energ\'{\i}a At\'omica, Av. Bustillo 9500, 8400 San Carlos de Bariloche, R\'{\i}o Negro, Argentina}
\author{Alejandro B. Kolton}
\affiliation{CONICET, Centro At\'omico Bariloche-Comisi\'on Nacional de Energ\'{\i}a At\'omica, Av. Bustillo 9500, 8400 San Carlos de Bariloche, R\'{\i}o Negro, Argentina}
\author{Leticia F. Cugliandolo}
\affiliation{Universit\'e Pierre et Marie Curie -- Paris VI, LPTHE UMR 7589,
4 Place Jussieu,  75252 Paris Cedex 05, France}
\date{\today}

\begin{abstract}
We study the thermally assisted relaxation of a directed elastic line in a two dimensional quenched random potential
by solving numerically the Edwards-Wilkinson equation and the Monte Carlo dynamics of a solid-on-solid lattice model.
We show that the aging dynamics is governed by a growing correlation length displaying two regimes: an initial
thermally dominated power-law growth which crosses over, at a static temperature-dependent correlation length $L_T \sim T^3$, to a logarithmic growth
consistent with an algebraic growth of barriers. We present a scaling arguments to deal
with the crossover-induced geometrical and dynamical effects. This analysis allows to explain why the results of
most numerical studies so far have been described with effective power-laws and also permits to determine
the observed anomalous temperature-dependence of the characteristic growth exponents.
We argue that a similar mechanism should be at work in other disordered systems.
We generalize the Family-Vicsek stationary scaling law to describe the roughness by incorporating the waiting-time dependence or age of the
initial configuration. The analysis of the two-time linear response and correlation functions shows
that a well-defined effective temperature exists in the power-law regime.  Finally, we discuss the
relevance of our results for the slow dynamics of vortex glasses in High-Tc superconductors.
\end{abstract}


\maketitle
%\tableofcontents

\section{Introduction}

Disordered elastic manifolds play an important role in a variety of physical
systems. Growing surfaces,~\cite{Barabasi-Stanley} interfaces in
domain growth phenomena and phase separation,~\cite{Alan} and cracks in
brittle materials~\cite{cracks} are usually viewed as elastic objects.
Interesting one dimensional realizations are
polymers~\cite{polymers} and vortex flux lines in type II superconductors.~\cite{rusos}  While
polymers can wrap, vortex flux lines are preferentially directed along the
applied magnetic field.

A number of systems involving one dimensional elastic manifolds
display glassy features.  An ensemble of interacting polymers forms a
polymer melt that undergoes a glass transition.~\cite{polymers} The
competition between vortex repulsion and their pinning to randomly
located impurities also leads to glassy phases in
superconductors.~\cite{rusos} While in the former case disorder is
self-induced, in the latter the effect of the impurities is mimicked
by a quenched random potential.

The relaxation dynamics of a model of layered high $T_c$
superconductors
was recently studied in Ref.~\onlinecite{us}.  The magnetic vortices are
considered to be directed along
one direction and they can move in two transverse
directions. Conventionally, this is a $(d\!=\!1)$+$(N\!=\!2)$
dimensional model. The dynamics following a rapid quench from the
liquid state to a set of control parameters in which the equilibrium
state is expected to be the so-called `vortex glass'~\cite{rusos} was
monitored.  By using numerical simulations
a slow out of equilibrium
relaxation typical of glasses was found.
In this system, diffusive aging of the averaged two-time roughness,
$\langle w^2\rangle$, and displacement, $\langle B \rangle $, was
observed. The aging was characterized by a {\it multiplicative}
scaling form $\ell^{2 \zeta}(t_w) f_{w^2,B}(\ell(t)/\ell(t_w))$, where
$\ell(t)$ is a growing characteristic length. The scaling functions
$f_{w^2}$ and $f_B$ are different but the exponent $\zeta$ does not
depend on the observable ($w^2$ or $B$).  Besides, the system of
interacting lines was perturbed in order to compute the integrated
linear response, and a diffusive aging was also found, characterized
by a scaling function of the type $\ell^{2 \zeta}(t_w)
f_\chi(\ell(t)/\ell(t_w))$ with the {\it same} exponent $\zeta$ and
growing length $\ell(t)$ found for the unperturbed two-time
observables (roughness and displacement).
It was also shown that correlation and response functions can be
related by a modified fluctuation-dissipation relation after removing
the `diffusive' contribution, {\it i.e.} the factors
$\ell^{2 \zeta}(t_w)$. This violation of the fluctuation-dissipation
relation can be characterized by a {\it finite} effective
temperature, $T_{\rm eff}$.~\cite{Cukupe,Leto}
Furthermore, within the time-window that was numerically explored, the
growing length $\ell(t)$ is well-described by a power-law, $\ell(t)
\sim t^{1/z}$.  This could well be a {\it pre-asymptotic} regime
after which the growing length crosses over to the expected
activated dynamics logarithmic growth (when disorder free-energy
barriers scale as $\Delta(L)\sim L^\psi$ with $\psi>0$).
It is worth to mention that although aging in high~\cite{high-Tc} and low~\cite{low-Tc}
$T_c$ superconductor samples has been reported in the literature, a detailed comparison to the results listed above remains to be done.

The model system used to describe vortex lines in high temperature superconductors contains several
contributions coming from different energy scales. The main contributions are:
the elastic energy of the individual lines,
non-linear terms in the elastic energy,
the interaction between the lines,
and the quenched random pinning potential.
It is clearly important to establish whether all the above listed
contributions to the energy are necessary to get such aging behavior or
whether similar features arise when some of the terms --
non-linear contributions to the elasticity, interactions between the
lines, random potential -- are switched off. We review below
the out of equilibrium dynamics of related models with and without this
type of interactions.

The first studies of the out of equilibrium dynamics of directed
elastic manifolds in a quenched random environment focused on {\it
  mean-field} models in which the transverse space is infinitely
dimensional, $N\to\infty$, each transverse coordinate has infinite
length, $M \to \infty$, and the manifold has finite dimension, $d$, and
infinite length in all directions, $L\to\infty$.~\cite{Cule}  This
model includes only two of the energetic contributions listed above
{\it viz.} elasticity and quenched disorder and neglects non-linear
terms and interactions between the elastic objects.  A dynamic phase transition
separating a liquid (high-temperature) phase from a glassy
(low-temperature) one was found for all internal dimension $d$
including $d=0$.~\cite{Cule}  Different aging dynamics characterize
the low-temperature phase depending on the short or long range
character of the random potential correlations. In both cases the
aging regime lasts for ever (after having taken the $N\to\infty$,
$M\to\infty$ and $L\to\infty$ limits): there is no
multiplicative factor $\ell^{2 \zeta}(t)$ in the averaged two-time
observables.  For short-range correlated potentials the displacement,
roughness and linear responses scale as
$f_{w^2,B,\chi}(\ell(t)/\ell(t_w))$ and a single finite effective
temperature exists.

Soon after Barrat~\cite{Barrat} and
Yoshino~\cite{Yoshino,Yoshino-unp} studied a solid-on-solid model of a
single directed elastic line relaxing with Monte Carlo dynamics on a
disordered substrate with one ($N=1$) transverse direction.  This
system models elasticity in a rather extreme way, using a hard constrain, and includes quenched
randomness.  There are no interaction between different lines in this model.
These authors focused on the very long length limit in an
effectively infinite box ($M \gg L\gg a$ with $a$ the lattice
spacing) and found that this lattice model has a similar averaged
dynamics to the one found later in the vortex glass~\cite{us} below a
crossover temperature. The relaxation is slow, and the global
displacement and linear response show non-trivial {\it diffusive}
aging with {\it multiplicative} scaling. The characteristic length
scale also appears to be a power-law of time within the numerical time
window and the exponents $z$ and $\alpha=\zeta/z$ are temperature and
disorder strength dependent with the same qualitative trend as in the
fully interacting case.  More recently, we focused on the analysis of the averaged
and fluctuating two-time roughness of such solid-on-solid lines with
{\it finite length}, $L<\infty$.~\cite{EPL} On the one hand, we found that
the aging regime stops and
crosses over to saturation of the two-time roughness and free diffusion
of the displacement at a characteristic value of the time delay,
$\Delta t\equiv t-t_w=t_x$, that grows with the length of the line and
smoothly depends on other parameters in the model.  The saturation of the
two-time roughness is well described by a generalization of the
equilibrium Family-Vicsek scaling.~\cite{Favi} On the other hand,
the two-time roughness fluctuations are highly non-trivial but
can be characterized with a relatively simple argument by a scaling
function.

The numerical solution to the Langevin equation for free lines in
$N=2$ transverse dimensions~\cite{us} and, especially, the full
analytic solution to the Langevin dynamics of finite Edwards-Wilkinson
(EW) elastic lines~\cite{EW} in one transverse
dimension~\cite{Sebastian,Pleimling} ($N=1$) had also been considered,
without taking into account interactions, non-linear terms and quenched randomness.
The results suggested that not even
disorder is necessary to obtain similar averaged and fluctuating aging
dynamics of fully relaxing quantities.  The aging scaling and
saturation phenomenon of the averaged two-time roughness follow the
same scaling laws as above with growing length $\ell(t)\sim t^{1/z}$,
and temperature-independent exponents $z=2$, $\zeta=1/2$ and
$\alpha=\zeta/z=1/4$.  However, the noise-averaged linear response is
stationary in this `quadratic' model. When it comes to analyzing
fluctuations other differences appear. The distribution of the
two-time roughness satisfies a similar scaling law as in the
disordered problem although with a rather different scaling function
and the linear response simply does not fluctuate.

The effect of non-linear terms has been considered by
Bustingorry~\cite{Sebastian2,Spaniards} who analyzed the relaxation
dynamics of `clean' finite length ($L<\infty$) Kardar-Parisi-Zhang
(KPZ) lines~\cite{KPZ} in $N=1$ transverse dimensions.
In this work correlations were analyzed in
detail and undergo diffusive aging with the KPZ
exponents $\zeta$, $\alpha$ and $z$.
The two-time and length scaling of averaged quantities and the
scaling form of the probability distribution functions proposed
in Ref.~\onlinecite{EPL} were thus confirmed.

Finally, once the elastic lines are allowed to wrap, biologically
motivated dynamic problems can also be addressed.  A recent study
concentrates on the effects of confinement on the out of equilibrium
relaxation of a single polymer chain in two dimensions,~\cite{claudio}
a problem of relevance for cellular modelling. Another study analyzes
numerically the effect of randomly applied forces on an ensemble of
interacting polymer lines, focusing on out of equilibrium properties
of active matter.~\cite{Davide} There is, certainly, much room
for further investigations in the biological context.

In this paper we return to the problem of the relaxation of directed elastic
lines in the presence of quenched randomness. Interactions among different
lines and non-linear terms are not considered here. Our aim is to complete
the analysis of the averaged two-time observables. In order to compare
with previous results and to extract the universal behaviour we
treat two models in parallel: the Monte Carlo dynamics of the
disordered solid-on-solid model~\cite{Barrat,Yoshino,Yoshino-unp} and
the Langevin dynamics of the disordered Edwards-Wilkinson equation in
$(1+1)$ dimensions. We present a careful study of finite-length
effects on the averaged dynamics of correlation
and linear response functions. We pay special attention to the
behavior of the crossover length between thermal and disorder dominated regimes,
and discuss finite-size and finite-time effects.
Moreover, we show by using scaling arguments how this length scale determines the
relaxation properties of the system.

The organization of the paper is the following. In
Sect.~\ref{sec:models} we define the models and we describe the
numerical method and two-time observables on which we focus. In
Sect.~\ref{sec:exponents} we measure the growth, roughness and dynamic
exponents by analyzing the two-time structure factor and two-time
roughness.  We also study in detail the growing length $\ell(t)$ and
we confirm the existence of a crossover from power-law to logarithmic
growth that is, however, only seen for sufficiently long lines.
Section~\ref{sec:averaged} summarizes the two-time behavior
of the averaged roughness and associated linear response using
different initial conditions.  In Sect.~\ref{sec:discusion} we
present scaling arguments to explain many of the features
observed numerically.
Finally, in Sect.~\ref{sec:conclusions} we present
our conclusions.

\section{\label{sec:models} The models}

In this section we introduce the models and the numerical
methods used to study their relaxation.

\subsection{Lattice model}

A discrete model of a one-dimensional directed elastic object
represents the line as a lattice string of length $L$ directed
along the $y$ direction.~\cite{Barrat,Yoshino,Yoshino-unp,EPL}  The
line can move transversely along the $x$ direction
on a rectangular square lattice of
transverse size $M=10^4\gg L^{2/3}$ ensuring the existence of many
nearly equivalent, quasi-ground states.  The line segments, $x(y)$
($y=1,\cdots,L$), obey the restricted solid-on-solid (SOS) rule
$|x(y)-x(y-1)|=0, 1$.
 A quenched random potential $V$ taking
independent values on each lattice site is drawn from a uniform
distribution in $[-1,1]$. We use $10^5-10^7$ realizations of the
quenched randomness depending on the value of $L$.  At each
microscopic time step we attempt a move of a randomly chosen segment
to one of its neighbors restricted by the SOS condition and we accept
it with the heat-bath rule.  One Monte Carlo (MS) step is defined as
$L$ update attempts. We use angular brackets to indicate the average
over thermal noise realizations. We choose two types of initial
conditions: equilibrium at high temperature obtained after evolving a
random initial condition during a sufficiently long time interval at
high temperature, and equilibrium at zero temperature.

The lattice model has no finite elastic
energy. Elasticity is modelled in an extreme way; in the absence of
disorder all configurations have equal vanishing energy but the
displacement between neighboring bonds is bounded to $-1,0,1$ lattice
spacings.

The control parameters are temperature, $T$, and the disorder
strength, $V_0=\sqrt{[V^2]_V}$.  Here and in what follows we use
square brackets, $[\dots]_V$, to indicate an average over quenched
randomness.  The Monte Carlo rule implies that the dynamics depend
only on the ratio between these parameters, $V_0/T$. In the
simulations shown here we fixed $V_0=1/3$.  In the following we use
adimensional time, space and energy scales.

\subsection{Continuous model}

The SOS model is easy to simulate but it is not simply recovered as a
limit of better-known continuous problems like the KPZ
equation~\cite{KPZ,Sebastian2} or the vortex line model~\cite{us} in
which elasticity is modelled in a more realistic way.  For this reason
we also study the  disordered
Edwards-Wilkinson (EW) line.~\cite{Barabasi-Stanley}

The disordered EW equation~\cite{EW} for a scalar
field $x$ representing the height of a surface over a one dimensional
substrate parametrized by the coordinate $y$ (a one dimensional
directed interface) is
\begin{eqnarray}
&&
\gamma \partial_t x(y,t) =
c \partial^2_y x(y,t)+F_p[x(y,t)]+h[x(y,t)]
\nonumber\\
&&
\qquad\qquad\;\;\;\;
+\xi(y,t),
\label{eq:EW}
\end{eqnarray}
where $\xi$ is a Gaussian thermal noise with
$\langle \xi(y,t) \rangle = 0$ and
\begin{eqnarray}
\langle \xi(y,t) \xi(y',t') \rangle =
2\gamma T \delta(y-y') \delta(t-t').
\end{eqnarray}
The parameter $c$ is the elastic constant, $\gamma$ is the friction
coefficient, $T$ the temperature of the thermal bath (in energy units)
and $\langle
\cdots \rangle$ the average over the white noise $\xi$, {\it i.e.} the
thermal average. The Boltzmann constant has been set to one, $k_B=1$.
The term $h$ represents the effect of a perturbing field that
couples linearly and locally to the height, $-h(y,t) x(y,t)$. One can
also consider other types of perturbation that couple to more
complicated functions of the height as defined in
Sect.~\ref{subsec:measurements}.
The random pinning force $F_p[x(y,t)] = - \partial_x V(x,y)$
represents the effect
of a random-bond disorder described by the potential $V(x,y)$, whose sample
to sample fluctuations are given by
\begin{equation}
\left[ \left[V(x,y)- V(x',y') \right]^2 \right]_V
= \delta(y-y') \, R(x-x')
\; ,
\label{eq:correlator}
\end{equation}
where $R(u)$ stands for a short ranged correlator along the $x$ coordinate
with range $r_f$. The continuous random potential is modeled by a cubic spline
passing through $M$ regularly spaced uncorrelated Gaussian number
points.~\cite{rosso_depinning_simulation} We adimensionalize
Eq.~\eqref{eq:EW} by using $r_f$ as the unit of distance in the $x$ direction,
$\gamma r_f$ as the unit of time, and $V_0$ as unit of energy/temperature. The
unit of distance in the longitudinal direction $y$ can be conveniently taken
as the layer spacing $s$ of the numerically discretized laplacian (such a
choice is natural when modeling a layered material such as a High-T$_c$
superconductor with an external magnetic field applied perpendicular to the
oxide planes). In these units, the friction and elastic coefficients are equal
to one, leaving only two independent parameters in the model: the
adimensionalized elastic constant $\nu = c r_f/s^2$ and the adimensionalized
temperature~$T/V_0$~ (from now on, we use $T$ for $T/V_0$). Finally, by
choosing $\nu=1$ we focus exclusively in the temperature dependence for a
fixed disorder.

We use a finite-difference algorithm to integrate the partial
differential equation in which the first and second order partial
derivatives are discretized in the usual way. (See Ref.~\onlinecite{Sebastian2}
for more details on the numerical technique.) We typically simulate
lines with length $L = 64, \ 256, \ 1024$. The time step is $t_0 =
0.01$.  We use $10^3$ noise
realizations to evaluate the two-times averaged correlations and responses.

Note that in the continuous model each non strictly flat configuration has a
finite elastic energy while in the lattice model all configurations have the
same vanishing elastic energy.

\subsection{\label{subsec:randomness} Quenched randomness}

The kind of quenched disorder used in both models is of the `random bond' type
in the sense that, for a domain wall described by our elastic interface
model, it does not break the symmetry properties of the corresponding
order parameter.~\cite{Nattermann} This implies that the two dimensional
disorder potential locally couples to the interface position and that $R(u)$
in Eq.~\eqref{eq:correlator} saturates to a constant for large distances. The
way we generate the disorder corresponds to a short-ranged function $R(u)$.
The effect of long-range correlated randomness has been studied in mean-field
elastic manifold models~\cite{Cule} but we do not contemplate it here.

\subsection{\label{subsec:measurements} Observables}

We study the relaxation after two types of rapid changes in the
control parameters. The first protocol, and the more usual one,
describes a quench from high to low temperatures. After equilibration
at high $T$ the system is quenched to a low value of
$T$. The second protocol, less usual, consists in first equilibrating
the sample at zero temperature ({\it i.e.} starting in its ground-state),
$T_0=0$, and then heating the sample
to a higher temperature. In both cases,
the line is allowed to relax from the quench
occuring at time $t=0$ until a waiting-time $t_w$, when
the quantities of interest are recorded and later compared
to their values at subsequents times $t>t_w$.

The aging dynamics of elastic lines has been initially studied in
terms of the averaged two-time
mean-squared-displacement $ \langle
B\rangle(t,t_w) = \left\langle \left[\left[ x(y,t) - x(y,t_w)
    \right]^2\right]_V \right\rangle $.~\cite{Barrat,Yoshino}
The behavior of this
quantity is somehow obscured by the motion of the center of mass.  We
thus prefer to focus on the roughness of the lines, a quantity that
has been widely used in the study of interface
dynamics,~\cite{Barabasi-Stanley} but generalized here to include the
$t$ and $t_w$ dependence. The two-time roughness is given by
\begin{equation}
\label{e:w2}
\langle w^2 \rangle (t,t_w)  = \frac{1}{L}
\sum_{y}
\left\langle \left[ \left[ \delta x(y,t) - \delta x(y,t_w) \right] ^2 \right]_V \right\rangle,
\end{equation}
where $\delta x(y,t)=x(y,t)-\overline{x}(t)$ accounts for the
displacement of the $y$-th line segment relative to the center of
mass, $\overline{x}(t)=L^{-1}\sum_y x(y,t)$.

The two-times structure factor is defined
as~\cite{Barabasi-Stanley,Sebastian}
\begin{eqnarray}
\langle S_n\rangle(t,t_w) &=& L \; \left\langle \left[ |c_n(t)-c_n(t_w)|^2 \right]_V \right\rangle
\; ,
\nonumber\\
L \; c_n(t) &=&  \sum_y \, [x(y,t)-\overline x(t)] \, e^{-i q_n y}
\;
\label{eq:structure-factor}
\end{eqnarray}
where
$q_n=2\pi n/L$ with $n$ integer.

Further insight into the
dynamics of out of equilibrium systems is given by the linear response
function. The latter is defined by applying a random time-independent force at
a time $t_w$ on a replica of the system, and by computing how this one departs
from an unperturbed one evolving with the {\it same} thermal noise.
Concretely, the energy contribution of a field  conjugated to the
displacement with respect to its mean value is
\begin{equation}
{\cal H}'^{w^2}=-h \;
\sum_y  \left[ x(y,t)- \overline{x}(t) \right]
s(y)
\theta(\Delta t)
\; .
\end{equation}
$s(y)=\pm 1$ with equal probability,
$\langle\langle s(y) \rangle\rangle=0$ and
$\langle\langle s(y) s(y') \rangle\rangle=\delta_{y,y'}$~\cite{Kolton-us}
with $\langle\langle \dots \rangle\rangle$ denoting the average
over the perturbing field distribution.
$h$ is the intensity of the perturbation.
The associated linear response function is
\begin{eqnarray}
\langle \chi\rangle(t,t_w) =
 \frac{1}{hL} \sum_y \; \left\langle
\left[ [\delta x^h(y,t) - \delta x(y,t)] s(y) \right]_V \right\rangle
\; .
\end{eqnarray}
Henceforth $\langle \ldots \rangle$ indicates the average over the
thermal noise {\it and} the $s(y)$
distribution.

In equilibrium the averaged linear response is related to the averaged
spontaneous fluctuations of the corresponding observable by the
model-independent fluctuation-dissipation theorem (FDT) which states
\begin{equation}
\langle w^2\rangle (\Delta t)=2T \langle \chi\rangle (\Delta t)
\end{equation}
(the Boltzmann constant has been set to one, $k_B=1$),
where the $\Delta t$ argument implies stationary dynamics. In a system
relaxing out of equilibrium this relation does not necessarily
hold. In a number of glassy systems one can
define an effective temperature~\cite{Cukupe} from the modification of the
above relation. In the aging
regime of elastic lines in disorder media
the FDT is violated and one constructs the modified
FDT~\cite{Leto,Yoshino,us}
\begin{equation}
\label{e:BXfdt}
\langle w^2\rangle (t,t_w)=2T_{\rm eff}(t,t_w) \;  \langle \chi\rangle(t,t_w).
\end{equation}
The two-time dependence of the effective temperature is
here kept general. It turns out that in models with
multiplicative scaling, as the one discussed here, once the factors
$\ell^\zeta(t)$ have been taken into account, the re-defined effective
temperature approaches a constant value (see Sect.~\ref{subsec:fdt}
and Refs.~\onlinecite{Barrat,Yoshino,Yoshino-unp,us,Sebastian,Kolton-us}).

\subsection{Dynamic crossover}

In $(1+1)$ dimensional models of elastic manifolds the glassy phenomenon
appears as a dynamic
crossover.~\cite{Barrat,Yoshino,Yoshino-unp,EPL,Sebastian,Pleimling,Sebastian2}

For all observation times, $t_{obs}$, that are longer than a size,
disorder-strength and temperature dependent {\it equilibration time},
$t_{eq}$, the system reaches equilibrium. Instead,
for $t_{obs} < t_{eq}$ the relaxation is non-stationary and
thus occurs out of equilibrium as demonstrated by
two-time correlations and linear responses that age and by a non-trivial
relation between them.

The equilibration time $t_{eq}$ increases
by increasing the size $L$ of the line, by
decreasing the temperature $T$, and by increasing the
quenched disorder strength $V_0$. At fixed disorder strength and line length the
dynamic crossover reflects in a temperature crossover that resembles a
phase transition at a characteristic temperature $T_{co}$.
At high temperature, $T>T_{co}$, the equilibration
time is short, disorder is irrelevant and the line behaves as the
clean EW one at high temperature: the dynamics of a finite length line
reaches equilibrium meaning that stationary dynamics and the FDT hold,
and the exponents (defined and discussed below) are the ones of
the clean EW line, $\alpha=1/4$, $\zeta=1/2$ and $z=2$.  At
lower temperatures, $T<T_{co}$, the equilibration time is longer than
the observation time and the dynamics remains non-stationary.~\cite{Barrat,Yoshino,Yoshino-unp,EPL}
Quenched disorder modifies the values of the exponents and induces a
geometrical crossover at a temperature dependent
length-scale $L_T$, from a short length-scale roughness
described by the `thermal' exponent $\zeta_T$, to a large length-scale
roughness described by a `disorder' exponent  $\zeta_D$.
We shall discuss the values of the exponents and the
out of equilibrium dynamics in the following sections.

\section{\label{sec:exponents} Growth and saturation}

In this Section we recall the main tools used to analyze the evolution
of the conformational properties of elastic manifolds and we
generalize them to take into account the effect of the waiting-time.
This leads us to present the temperature dependence of the crossover length scale $L_T$.

\subsection{Roughness}

\subsubsection{Comparison to the initial condition}

Traditionally, the dynamics of elastic manifolds has been classified
in universality classes according to the behavior of the two-time
roughness \eqref{e:w2} evaluated at $t_w=0$, the initial time, and
using a flat initial configuration,
$x(y,0)=x_0$.~\cite{Barabasi-Stanley,Favi}  The waiting-time being
identical to zero all two-time observables depend on $\Delta t=t$ when
compared to the initial condition.  Initially the roughness increases
as a function of $\Delta t$, and at a characteristic time
$t_x(L)$ reaches saturation at an $L$-dependent value, $\langle
w_\infty^2\rangle$.  This behavior is encoded in the Family-Vicsek
scaling~\cite{Favi} that, in full generality, can be expressed as
\begin{eqnarray}
\langle w^2\rangle(\Delta t) &\sim& g(\Delta t)
\; , \label{eq:time-gen} \\ t_x &\sim& f(L) \; ,
 \label{eq:cross-overtime-gen}
\\
\langle w^2_\infty\rangle \equiv \lim_{\Delta t\gg t_x} \langle w^2\rangle
(\Delta t)
 &\sim&
 h(L) \; ,
\label{eq:saturation-gen}
\end{eqnarray}
 with
$h,g,f$ three monotonic functions. Consistency at $\Delta t=t_x$ requires $h=f
\circ g$.  In the usually discussed space-time scale-invariant cases in which all functions
are power-laws one has
\begin{eqnarray}
\langle w^2\rangle(\Delta t) &\sim& \Delta t^{2 \alpha}
\; ,
\label{eq:time}
\\
t_x &\sim& L^z
\; ,
\label{eq:cross-overtime}
\\
\langle w^2_\infty\rangle &\sim& L^{2 \zeta}
\; ,
\label{eq:saturation}
\end{eqnarray}
with
$\alpha$ the growth exponent, $z$ the dynamic exponent,
and $\zeta$ the roughness exponent.
Consistency implies that the three exponents are related by
\begin{equation}
z\alpha =\zeta \; .
\label{eq:exponents}
\end{equation}
The values of the exponents are well-known in a number of cases. For
the EW elastic line the exponents can be analytically computed and one
finds that $\alpha=(2-d)/4$, $z=2$ and
$\zeta=(2-d)/2$ for $d \leq 2$. For the non-linear KPZ elastic line the exponents
are known only numerically for general
$d$. In $(1+1)$ dimensions, however, they can be computed analytically
and $\alpha=1/3$, $z=3/2$ and $\zeta=1/2$.

In the presence of quenched disorder the dependence of the asymptotic
roughness $\langle w^2_\infty\rangle$ with the length of the line
undergoes a crossover. For lines that are shorter than a temperature
and disorder strength dependent value $L_T$ the behavior is
controlled by thermal fluctuations and the relation
\eqref{eq:saturation} holds with $\zeta=\zeta_T$ the thermal roughness
exponent. This exponent is the one corresponding to the EW equation, and thus
$\zeta_T=(2-d)/2$ in general, and $\zeta_T=1/2$ in our $(1+1)$
dimensional case.  In this thermally dominated scale, the dynamics is expected to be `normal' in the sense that
lengths and times should be thus related by power-laws of the type
\eqref{eq:time}-\eqref{eq:saturation} with the
exponents linked by Eq.~\eqref{eq:exponents}.

For lines longer than $L_T$, the roughness is dominated by
quenched disorder and one has that \eqref{eq:saturation} still
holds though with a different value of the roughness exponent,
$\zeta=\zeta_D$. The disorder dominated roughness exponent $\zeta_D$ is the one characterizing the geometry of the
ground state configurations, and it is expected to
satisfy $\zeta_D>\zeta_T$ in general.~\cite{Kardar}
In $(1+1)$ dimensions one has $\zeta_D=2/3$.

Once quenched disorder is present, the time evolution is expected
to be driven by activation over free-energy barriers. If they
scale as $U(L) \sim L^\psi$, the Arrhenius law leads to a
logarithmic relation between equilibrated
lengths and times that implies
$t(\ell) \sim e^{\Upsilon (\ell/L_\Upsilon)^\psi/T}$ with $\Upsilon$ and $L_\Upsilon$ some characteristic energy and length scale, respectively.
One might then expect
\begin{eqnarray}
\langle w^2\rangle(\Delta t) \sim \ln^{\alpha_{\ln}} \Delta t,
\end{eqnarray}
which for consistency implies
\begin{equation}
\psi \ {\alpha_{\ln}}= 2\zeta_D
\; .
\label{eq:psi-alpha-zeta}
\end{equation}

All exponents cited above are temperature independent.
Some glassy systems do, however,
present temperature-dependent exponents
asymptotically. Whether this behavior can occur in our
system is a delicate issue, that we discuss in
Sect.~\ref{sec:discusion}.

\subsubsection{Comparison to an aged configuration}

In simple cases in which {\it equilibrium} is relatively rapidly
reached the initial condition should be irrelevant
after the equilibration time.  The same relaxational behavior is then expected,
independently of the waiting-time $t_w$ and the initial condition. The
roughness should only depend on the time-difference $\Delta t=t-t_w$
and the same functional forms should characterize growth and
saturation.
In an {\it out of equilibrium} relaxation relatively soon after preparation
the growth regime acquires a
waiting-time dependence and the relations
\eqref{eq:time-gen}-\eqref{eq:saturation-gen}
might be generalized.

For each waiting-time the two-time roughness presents two regimes as a
function of $\Delta t$: it first grows until crossing over at $t_x(L)$
to saturation at a $\Delta t$ independent value.
For high working temperature after a short transient
the memory of the initial  condition
disappears, and the roughness is well-described by the Family-Vicsek
scaling. For low working temperatures the waiting-time dependence remains.
For long lines the crossover time is
long enough to see aging behavior in a sufficiently long
time-window, such that we can describe it with a scaling
form. Before saturation the roughness does not
 depend strongly on the size of the line (for sufficiently long
lines) and  the curves can be scaled as
\begin{equation}
\langle w^2 \rangle(t,t_w)
\sim
\ell^{2 \zeta(t_w)} \;
\langle \tilde w^2\rangle
\left(\frac{\ell(t)}{\ell(t_w)}\right)
\; .
\label{eq:scaling}
\end{equation}
This scaling form approaches a stationary regime
in which $\langle w^2\rangle(t,t_w) \sim \ell^{2 \zeta}(\Delta t)$
in the limit $\Delta t \gg t_w$ if $\langle \tilde w^2\rangle
(u)\sim u^{2 \zeta}$ for $u\gg 1$. For even longer time-delays such
that $\ell(\Delta t) \to L$ one finds saturation at
$\langle w^2\rangle(t,t_w) \to \langle w_\infty^2\rangle \sim L^{2 \zeta}$,
which might indeed contain a $t_w$-dependent prefactor,
as in the clean EW case.~\cite{Sebastian}

The out of equilibrium regime exists without the need of quenched
randomness. We showed in Ref.~\onlinecite{Sebastian} that the roughness in the
simple EW equation satisfies the scaling form \eqref{eq:scaling} when
the system is let evolve from an out of equilibrium initial condition
and the waiting-time dependence is kept explicit. The length
scale $\ell$ grows as $\ell(t) \sim t^{1/2}$. A similar scaling was
shown numerically in Ref.\onlinecite{Sebastian2} for the $(1+1)$ KPZ model with
$\ell(t)\sim t^{2/3}$.  In Sect.~\ref{sec:averaged} we discuss the
scaling of Eq.~\eqref{eq:scaling} for the disordered model.

The analysis of the saturation of the two-time roughness should, in principle,
yield the thermal and disorder-dominated values of $\zeta$.
Computing $\zeta$ from $\langle w^2_\infty \rangle$ is, however,
numerically heavy, since one needs to simulate lines with different lengths
and then extract the scaling behavior. In Ref.~\onlinecite{EPL} we tried
such a scaling analysis but we did not reach the regime in which
$\zeta=\zeta_D=2/3$ (see Fig.~1(d) in this reference).
A more convenient way of getting $\zeta$ is to study the two-time
structure factor,~\cite{Barabasi-Stanley} as we explain below.

\subsection{Structure factor}

\subsubsection{The roughness exponents}

\begin{figure}
\centerline{
\includegraphics[angle=-0,width=\linewidth,clip=true]{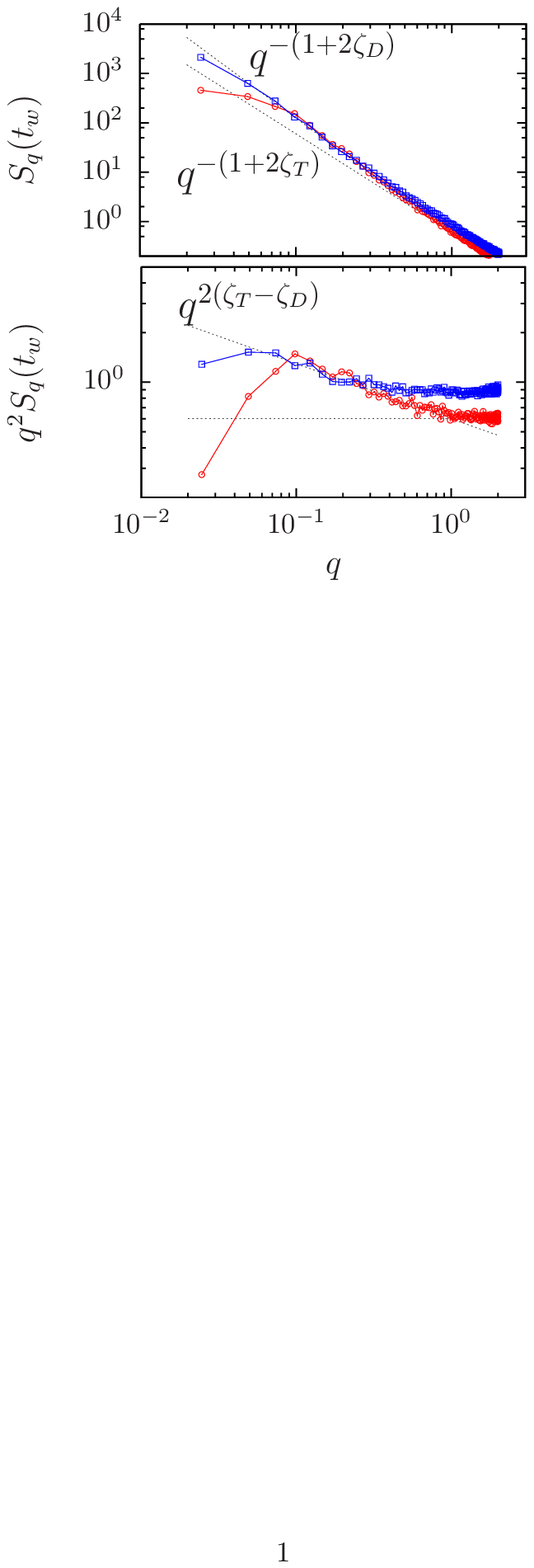}
}
\caption{(Color online) Structure factor in the continuous EW line in
  a random environment for $T_0=5$ and two working temperatures,
  $T=0.5$ (circles) and $T=0.8$ (squares), with $t_w=0$. The
  data show the thermal-regime at very large wave vector, a crossover
  to a disorder-dominated regime upon decreasing the wave-vector value
  and, finally, a saturation regime demonstrating that the
  growth-length, $\ell(t)$, is shorter than the system size, $L$.
  In the bottom panel the structure factor is scaled by $q^2$ in order
  to highlight the difference between the two roughness exponents.}
\label{fig:pru_1cTf0.5}
\end{figure}

The thermal, $\zeta_T$, and disorder, $\zeta_D$, roughness exponents can also
be extracted from the analysis of the structure factor defined in
Eq.~\eqref{eq:structure-factor}.  Working with
the structure factor is convenient since it is sufficient to simulate
a long chain and then extract the $L$ dependence from the
wave-vector dependence.~\cite{Barabasi-Stanley}
Indeed, the roughness is simply related to the structure factor as
\begin{equation}
\langle w^2\rangle(t,t_w) = L^{-1} \sum_{n=-\infty}^\infty \langle S_n\rangle(t,t_w),
\end{equation}
and this implies that
the dynamic scaling \eqref{eq:time-gen}-\eqref{eq:saturation-gen}
(for $t_w=0$)
is equivalent to
\begin{equation}
\langle S_n\rangle(\Delta t) \equiv \langle S_n\rangle(t=\Delta t, t_w=0) = S^{eq}(q_n) \; g[q_n \ell(\Delta t)]
\label{eq:scaling-Sq}
\end{equation}
with $S^{eq}$ the equilibrium wave-vector dependent structure factor
and $g(u) \sim u^{d+2 \zeta}$ for $u\ll 1$ and $g(u)\to \mbox{const}$
for $u\gg 1$, with $d$ the internal dimension of the manifold ($d=1$
for the directed line).
These limits for the function $g(u)$ ensure, respectively, the equilibrated steady geometry and the memory of the flat initial condition.
In the case of power-law scaling Eq.~\eqref{eq:scaling-Sq}
becomes
\begin{equation}
  \langle S_n\rangle(\Delta t) = q_n^{-(d+2\zeta)}\; g[q_n \Delta t^{1/z}]
  \; .
\end{equation}

In the out of equilibrium relaxation process we are interested,
$\langle S_n\rangle$ depends on two-times $t$ and $t_w$.  The
generalized scaling form then reads
\begin{equation}
\langle S_n\rangle(t,t_w) = S^{eq}(q_n) \; G[q_n \ell(t), q_n \ell(t_w)]
\end{equation}
that reduces to Eq.~\eqref{eq:scaling-Sq} both for $t_w=0$,
and to a similar form for $q_n \ell(t_w)
\gg 1$ for all $n$, and thus stationarity is recovered.
For fixed $t_w$ and $\Delta t$ we can study the dependence of
$\langle S_n\rangle$ on $q_n$. If the aim is to determine the values of the
roughness exponent the ideal choice would be to use a very long
$\Delta t\equiv t-t_w\gg t_w$ and plot $\langle S_n\rangle$ for
several $t_w$'s as a function of $q_n$. This construction is to be
compared with Fig.~2(b) in Ref.~\onlinecite{Sebastian}. In presence of
disorder the construction should have a broken straight-line form with
two exponents, $-(1+2 \zeta_{T,D})$; the thermal one, $\zeta_T$,
characterizing the behavior at short length-scales, $q_n>q_T$, and the
disorder one, $\zeta_D$, characterizing the behavior at long
length-scales, $q_n<q_T$.  The breaking point $q_T$ defines a characteristic length $L_T \sim 1/q_T$ which should depend on $T$ and we shall discuss and evaluate its precise
temperature dependence in Sect.~\ref{sec:crossoverlength}. In an infinite
$(1+1)$-dimensional system disorder always dominates the fluctuations
at very long length scales. However, in a finite system of length
$L$ once $L_T = L$ all the system is characterized by thermal
fluctuations and the value of $L_T$ is thus crucial.

Studying the wave-vector dependence of the structure factor 
is thus a convenient way to determine the
roughness exponents,
$\zeta_{T,D}$; we hence expect to improve the values shown in Fig.~1(d)
in Ref.~\onlinecite{EPL}.  We studied the $t_w=0$ structure factor in the
continuous disordered EW line.  The general behavior described in the
previous paragraphs is shown in Fig.~\ref{fig:pru_1cTf0.5} for
working temperatures $T=0.5$ and $T=0.8$.  Note that the saturation of the
structure factor at very small wave vector bears the same information
as the fact that the roughness $\langle w^2 \rangle$ has not yet
saturated at the longest time shown in Fig.~\ref{fig:pru_1cTf0.5}. It
means that the growing correlation length has not reached the size of
the system, $\ell(t) <L$.  The crossover from the thermal, $\zeta_T$,
to the disorder, $\zeta_D$, roughness exponents is clear,
with the expected values $\zeta_T=1/2$ and $\zeta_D=2/3$ confirmed.
Note that the top panel in Fig.~\ref{fig:pru_1cTf0.5} shows the
raw data while the bottom panel shows the structure factor multiplied by $q^2$,
which allows to better observe the crossover region and the difference
between the two exponents.

\subsubsection{\label{sec:crossoverlength} The crossover length scale $L_T$}

\begin{figure}
\centerline{
\includegraphics[angle=-0,width=\linewidth,clip=true]{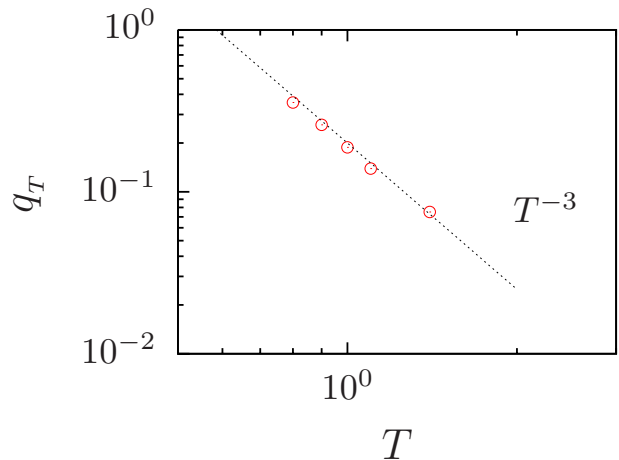}
}
\caption{(Color online) The value of the wave vector $q_T$ signaling
  the crossover between the thermal and disorder regimes at different
  working temperatures. The temperature dependence is consistent with
  the expected $q_T \sim T^{-3}$ power-law, see Eq.~\eqref{eq:phi}.}
\label{fig:pru_qc}
\end{figure}

The analysis of the structure factor also allows us to evaluate the
temperature and disorder strength dependence of the crossover length
$L_T$ (or wave-vector $q_T$)
and compare it to Yoshino's unpublished results for the lattice
model~\cite{Yoshino-unp} and Nattermann, Shapir and Vilfan's
estimate.~\cite{Nattermann}

The temperature-dependent crossover length $L_T$ separating the
saturation regime characterized by the thermal and disorder exponents
$\zeta_T$ and $\zeta_D$ should be observable in this presentation as
a temperature-dependent crossover wave-vector $q_T$ in the
equilibration prefactor $S^{eq}$. In order to access it one then needs
to use a sufficiently long time-delay in such a way that the second
factor saturates to $g(u)\to \mbox{const}$.

The crossover length, and the crossover wave-vector,
are expected to
scale with temperature as $L_T \sim T^{1/\phi}$, and $q_T\sim
T^{-1/\phi}$, with $\phi$ an exponent that we shall obtain below using
a variety of arguments. The characteristic length-scale $L_T \sim
1/q_T$ should increase with increasing temperature ($\phi > 0$),
indicating that
thermal fluctuations become more and more important.

A simple argument to obtain the $\phi$ exponent relies on the fact
that the fluctuations at large length scales ($q_n <q_T$) are not
expected to strongly depend on the working temperature, in contrast
with the short length scale ones. This
is indeed confirmed in our numerical simulations, as can be
observed comparing the two sets of data in Fig.~\ref{fig:pru_1cTf0.5}.
We will use this observation as a main input for the scaling argument
in Sect.~\ref{sec:discusion}.
Let us assume that a weak temperature dependence is permitted,
{\it i.e.} $\langle S_q \rangle
\sim a(T) q_n^{-(1+2 \zeta_D)}$ for $q_n<q_T$, while $\langle S_q \rangle \sim T
q_n^{-(1+2 \zeta_T)}$ for $q_n>q_T$. Matching the crossover between these two
regimes at $q_T$ one obtains
\begin{equation}
q_T \sim \left[ \frac{T}{a(T)} \right]^{-1/[2(\zeta_D-\zeta_T)]}
\; .
\label{eq:phi}
\end{equation}
Using $a(T) \sim a$ yields $q_T \sim T^{-1/[2(\zeta_D-\zeta_T)]}$
and $\phi=2(\zeta_D-\zeta_T)$.  With simulations of the discrete model
Yoshino found $\phi=1/3$,~\cite{Yoshino-unp} in agreement with this
result since for our one dimensional case $\phi=2(\zeta_D-\zeta_T)=2(2/3-1/2)=1/3$.

A different prediction for the exponent $\phi$ comes from an order of
magnitude argument. If one assumes that the characteristic {\it
  free-energy barrier}, $\Delta(L)$, associated to the length-scale
$L_T$ should be the thermal one,
\begin{equation}
\Upsilon \left( \frac{L_T}{L_0} \right)^{\psi} \sim k_B T
\; ,
\label{eq:thermal-scale}
\end{equation}
one finds
\begin{equation}
L_T\sim L_0 \left( \frac{k_B T}{\Upsilon} \right)^{1/\psi}
\; .
\label{eq:Lc-thermal-scale}
\end{equation}
If $\Upsilon$ and $L_0$ are independent of temperature, $L_T \sim T^{1/\psi}$
and $\phi=\psi$. Yoshino used this kind of argument
with the free energy barrier replaced by the sample-to-sample fluctuations
of the energy, an assumption tested in Ref.~\onlinecite{drossel} for low temperatures.
In other words he replaced $\psi$ by $\theta$, with $\theta$ the
energy exponent to get $\phi=\psi \approx \theta$ ($\theta=1/3$ in $d=1$).~\cite{Yoshino-unp}
This argument assumes that the scale $L_T$
is the onset for thermally activated motion,
making a connection between the dynamics and the geometry of the line.

On the other hand, using Flory-type arguments to predict the roughness
exponent, $\zeta_F=[5-(d+N)]/5=3/5$ for $d=1$, with $\langle w^2\rangle \sim
L^{2 \zeta_F}$, and an expansion around $T=0$, Nattermann {\it et
  al}~\cite{Nattermann} suggested $\phi\simeq 2(\zeta_F - \zeta_T)$ and
using $\zeta_T=[3-(d+N)]/2=1/2$ they found $\phi \simeq 1/5$
in $d=N=1$.

Figure~\ref{fig:pru_qc} displays the result of the analysis of the
crossover between the thermal and the disorder regimes in the $t_w=0$
structure factor at different working temperatures.
Our numerical simulations of the disordered EW
line is compatible with $\phi=1/3$ in
agreement with the first two exposed arguments and Ref.~\onlinecite{Yoshino-unp},
see Eq.~\eqref{eq:phi}, but it rules out
the value given by Nattermann {\it et al}.~\cite{Nattermann} However,
by replacing the Flory approximation to the roughness exponent
used by these authors by the exact exponent, we get $\phi=1/3$ again, in
agreenment with our results.

\subsubsection{\label{subsec:log-growth} The growing length: logarithmic growth from $t_w=0$ measurements}

At high temperature, $T>T_{co}$, equilibrium is quickly reached,
disorder is irrelevant and one recovers stationary dynamics with the
clean EW temperatue-independent values $\alpha=1/4$, $\zeta=\zeta_T=1/2$ and
$z=2$. At low temperature one has to distinguish between the thermal and
disorder regimes and analyze the width of the crossover region.

\begin{figure}
\centerline{
\includegraphics[angle=-0,width=\linewidth,clip=true]{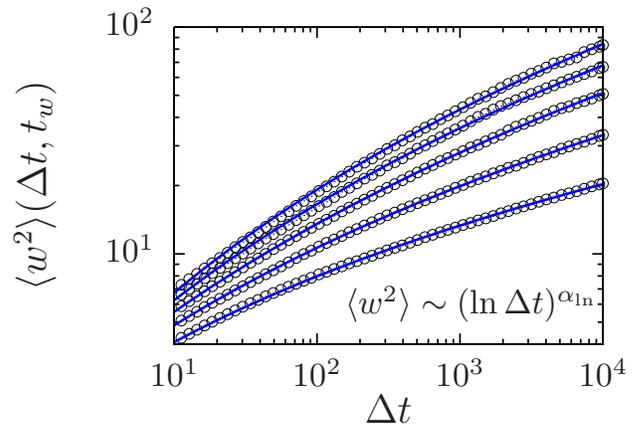}
}
\caption{(Color online) Logarithmic fit of the roughness relaxation of the
  elastic string with $t_w=0$. All the curves correspond to the same
  initial temperature $T_0=5$ and different quench temperatures
  $T=0.5,\,0.7,\,0.9,\,1.1$ and $1.3$ from lower to upper curves.
  Circles are raw data and blue continuous curves are the fitted
  functions.}
\label{fig:w2-ln_DT-alpha}
\end{figure}

\begin{figure}
\centerline{
\includegraphics[angle=-0,width=\linewidth,clip=true]{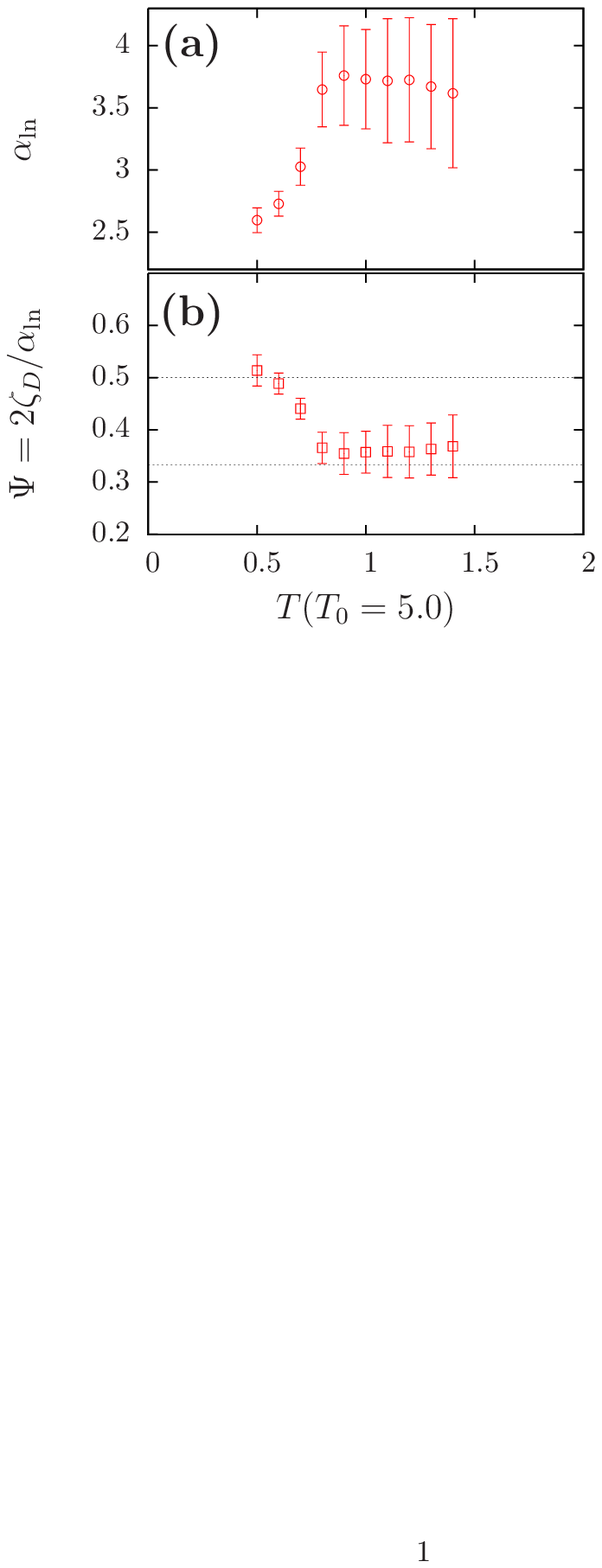}
}
\caption{(Color online) Dependence on the working temperature of
(a) the $\alpha_{\ln}$ exponent obtained from
  the fits in Fig~\ref{fig:w2-ln_DT-alpha}, and (b) the
  barrier exponent $\psi=2 \zeta_D/\alpha_{\ln}$, see
  Eq.~\eqref{eq:psi-alpha-zeta}, with the dotted lines indicating
  the suggested values $\psi =1/2$ and $\psi \approx \theta = 1/3$.
  The temperature of the initial condition is $T_0=5.0$. The error
  bars were estimated from the dispersion of the exponents when changing
  the fitting range.}
\label{fig:theta-fit}
\end{figure}

\begin{figure}
\centerline{
\includegraphics[angle=-0,width=\linewidth,clip=true]{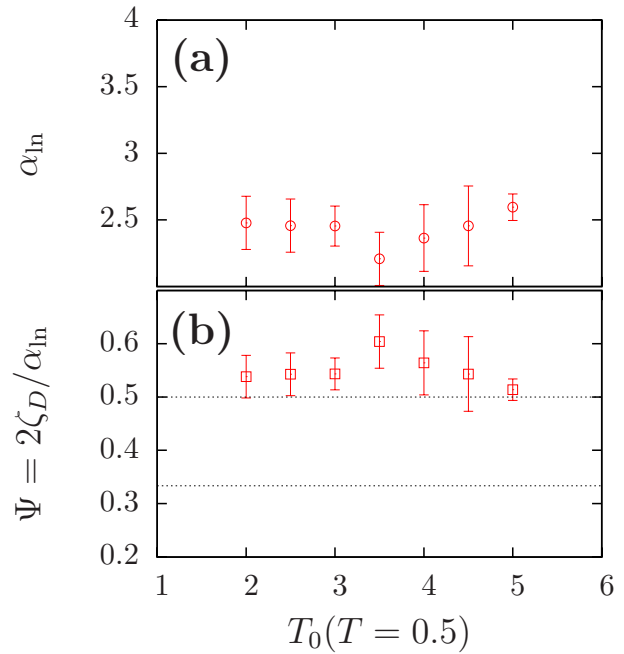}
}
\caption{(Color online) Exponents obtained from a logarithmic fit of
  the roughness relaxation of the elastic string for fixed working temperature
  $T=0.5$ and different initial conditions $T_0$. (a) $\alpha_{\ln}$
  exponent and (b) barrier exponent, $ \psi$. The dotted line
  indicates the suggested $\psi = 1/2$ and $\psi =1/3$ values. The error
  bars were estimated from the dispersion of the exponents when changing
  the fitting range.}
\label{fig:theta-fit-c}
\end{figure}

We shall first take $t_w = 0$ and compare with the results obtained by
Kolton {\it et al}~\cite{Kolton} (see also Ref.~\onlinecite{Yosh07}).
These authors showed, by studying the evolution of the structure factor of an elastic line
in random media from a flat initial condition and using $t_w=0$, that the characteristic
growing length crosses over from a power-law dependence $\ell(t) \sim
t^{1/2}$ typical of the clean EW case to a logarithmic law, $\ell(t)
\sim (\ln t)^{1/\psi}$ with $\psi\sim 0.49$ (see Fig.~2~in
Ref.~\onlinecite{Kolton}). The exponent $\psi$ characterizes the scaling of
the barriers at large scales, $\Delta(L)\sim
L^\psi$. Kolton {\it et al} expected $\psi=\theta$, with $\theta$ the
exponent characterizing the free-energy cost of an excitation of
length $L$, $\Delta F(L)\sim L^\theta$.  The value of $\theta$ is
known exactly for $d=N=1$, $\theta=1/3$, and the authors of Ref.~\onlinecite{Kolton} ascribed the discrepancy between the measured growing
length, $\ell(t)\sim (\ln t)^{1/0.49}$, and their expectation, $\ell(t)
\sim \ln^3 t$ to strong logarithmic corrections.

If the structure factor scales as
in Eq.~\eqref{eq:scaling-Sq}, the roughness scales as
\begin{eqnarray}
\langle w^2 \rangle(\Delta t)
\sim \int dq_n \, q_n^{-(1+2 \zeta)} \, g[q_n \, \ell(\Delta t)]
\sim \ell^{2 \zeta}(\Delta t)
\; .
\end{eqnarray}
Using a long line, $L=256$, as in Ref.~\onlinecite{Kolton} we
confirm the logarithmic growth of the roughness for the $t_w = 0$
case, as shown in Fig.~\ref{fig:w2-ln_DT-alpha}. Different curves
correspond to increasing temperatures, $T=0.5,\,0.6,\,0.7,...1.4$ from
bottom to top. We fit these curves to $\langle w^2 \rangle (\Delta t)=
A [\ln (\Delta t/B)]^{\alpha_{\ln}}$ obtaining the temperature
dependent $\alpha_{\ln}(T)$ exponent shown in
Fig.~\ref{fig:theta-fit}(a). Now, by simply using the relation $\langle
w^2\rangle(\Delta t) \sim \ell^{2 \zeta}(\Delta t) \sim
(\ln \Delta t)^{\alpha_{\ln}}$ one recovers the exponent $\psi$, {\it
  i.e.}  $\psi = 2 \zeta_D/\alpha_{\ln}(T)$, where we assumed that in
this regime the disorder roughness exponent is the relevant one.  We
obtain $\psi(T=0.5) \approx 0.51$ which is close to the value $\psi
\approx 0.49$ obtained in Ref.~\onlinecite{Kolton}.

Recently, Monthus and
Garel~\cite{Monthus} conjectured $\psi=d_s/2$ with $d_s$ the
dimension of the surface of the excitation, $d_s=1$ in our case, that
is in very good agreement with our numerical results at $T=0.5$
without any need to advocate for logarithmic corrections.
However, the analysis at different working temperatures presented here
reveals that the value of $\psi$ departs from $1/2$ and
approaches $1/3$ for increasing temperature, see Fig.~\ref{fig:theta-fit}(b).
This is the contrary to what is expected from the argument
given in Ref.~\onlinecite{Monthus} that
is based upon the role of the entropic contribution. Indeed, one would have expected
the value $1/2$ to prevail at high temperatures, which is the opposite
trend to what we find in the numerical simulations. We shall show in
Sect.~\ref{sec:discusion} that the temperature dependence of $\psi$ observed in
Fig.~\ref{fig:theta-fit}(b) can be explained as a finite size/time effect
induced by the crossover.

We have also analyzed the behavior of these exponents when changing
the initial temperature, while keeping $t_w=0$, that is to say, using
initial conditions thermalized at different $T_0$'s.  We performed the
same fitting procedure keeping the working temperature fixed to
$T=0.5$ and varying $T_0=2.0,\,2.5,\,3.0,\,3.5,\,4.0,\,4.5,\,5.0$. The
resulting exponents are shown in Fig.~\ref{fig:theta-fit-c}. Despite
the rather large fluctuations  we can trust that
there is no systematic dependence on the initial temperature. We
conclude that the exponent values do not depend on the previous
history but they do on the relaxation conditions.

\subsubsection{The growing length: pre-asymptotic power-law at finite $t_w$}

When the waiting time dependence is considered, the logarithmic growth
is no longer easily observed. On the one hand, the crossover between
the two asymptotic limits, $t_w \ll \Delta t$ and $t_w \gg \Delta t$,
makes it difficult to fit a logarithmic growth. On the other hand,
when the correlation length $\ell$ increases the necessary time to
reach the logarithmic regime increases exponentially $\tau \sim
\exp(\ell^\psi)$. Thus, from a practical point of view, the
logarithmic regime is not reached and one can use a power-law growth
of the roughness as an effective description. We shall use such a
power-law to analyze the aging behavior of the roughness in
Sect.~\ref{sec:roughness}.

In this case, the best option to extract the dynamic growing length
$\ell(t)$ is to take a very long $t_w$ and study the breaking
point $q_x \sim 1/\ell$ in the structure factor.
This breaking point separates the modes keeping the memory of the initial
condition, $q \ll q_x$, from the equilibrated modes described by the
power-law regime with disorder exponent, $q \gg q_x$.
In the case of a power-law scaling,
$t\sim L^z$, one has
\begin{equation}
q_x \sim \Delta t^{-1/z}
\end{equation}
and from here one obtains $z$.
Using power-law fits we find that the growth and
dynamic exponents, $\alpha$ and $z$, acquire a $T$-dependence
in the presence of quenched disorder~\cite{Barrat,Yoshino,Yoshino-unp,EPL}
while the roughness exponents
$\zeta_{T,D}$ do not depend on temperature.

\section{\label{sec:averaged} Averaged aging dynamics}

In this Section we analyze the two-time evolution of quantities that
are averaged over many realizations of the thermal noise and quenched
disorder. We consider two types of preparation: cooling from
a higher temperature and heating from zero temperature. We discuss
how the initial condition affects the behavior in the pre-asymptotic
growth regime.

\subsection{\label{sec:roughness} The two-time roughness}

The analysis of the averaged two-times roughness of a very long line
described by the solid-on-solid model following a quench from infinite
temperature was given in Refs.~\onlinecite{Barrat,Yoshino,Yoshino-unp,EPL}.  We
summarize here these results and we complement
them by showing that:
{\it i.} a similar scaling form describes the relaxation dynamics of
the disordered EW line; {\it ii.} the dynamics after heating up the
lines can also be described by the same scaling form with parameters
that depend on the working {\it and} initial temperature; {\it iii.}
the qualitative behavior is similar to the one found analytically for the
clean EW~\cite{Sebastian,Pleimling} and numerically for the clean
KPZ~\cite{Sebastian2} lines.

\begin{figure}
\includegraphics[angle=0,width=.85\linewidth,clip=true]{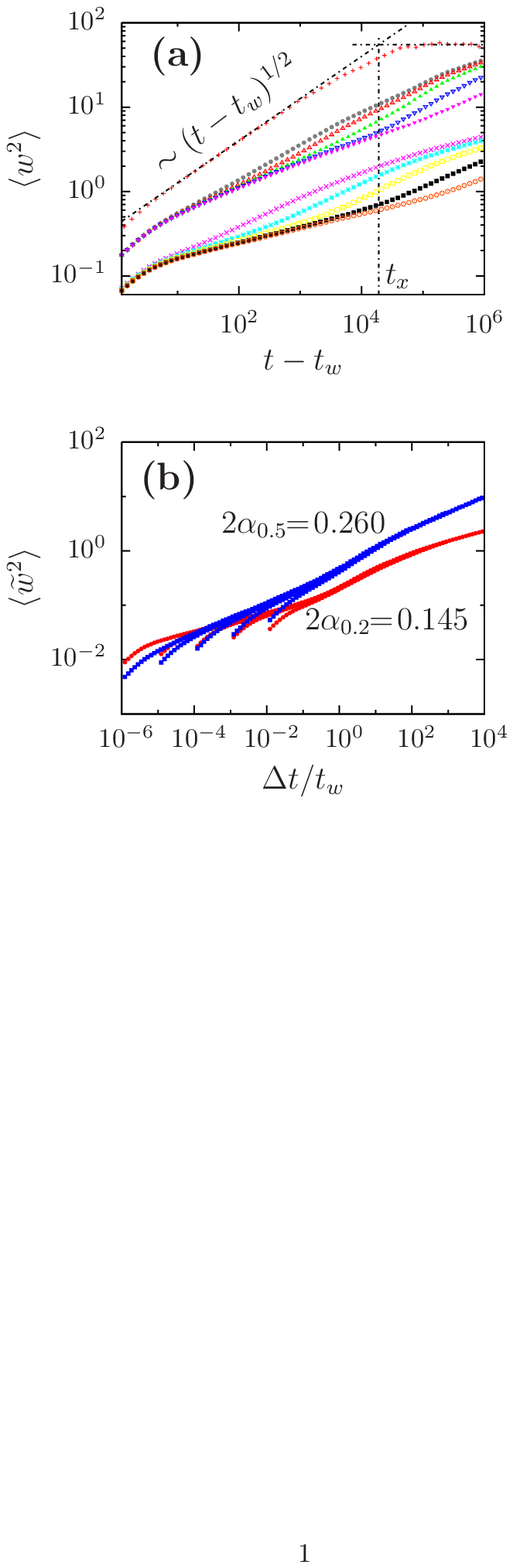}
\caption{(Color online)
Panel (a): the noise and disorder averaged two-time
roughness of the lattice elastic line with length $L=500$ evolving at
$T=5, \ 0.5, \ 0.2$ after a quench from infinite temperature, $T_0
\to\infty$.  Panel (b): scaling plot of the averaged two-time
roughness of the line with $L=500$ and $T=0.2$, and the line with
$L=5000$ and $T=0.5$.
}
\label{fig:averaged-roughness}
\end{figure}

The time-difference dependence of the averaged roughness evolving from
an infinite temperature initial condition in the SOS model,
for several waiting-times and at three working temperatures, is shown in
Fig.~\ref{fig:averaged-roughness}(a).  The averaged roughness of the
line evolving at high temperature, $T=5$, is stationary. It grows as
the clean EW roughness, $\langle w^2\rangle \sim \Delta t^{1/2}$ and
it reaches saturation at $\langle w_\infty^2\rangle$ after $\Delta
t>t_x$. The line with $L=500$ evolving at $T=0.5$ is not stationary
but signs of saturation are visible at $\Delta t \sim 10^6$ MCs.  The
line evolving at low temperature, $T=0.2$, shows aging effects, is
still in the growth regime and does not reach saturation in the
numerical time-window.

The averaged roughness curves in the growth regime can be well
described by the scaling in Eq.~\eqref{eq:scaling}.  The numerical analysis
of the growing length $\ell(t)$ in the time-window we can reach
indicates that $\ell(t)\sim t^{1/z}$. As discussed in
Sect.~\ref{subsec:log-growth} this power-law regime should be
considered as a pre-asymptotic approximation to a slower, logarithmic
growth.  We shall continue our analysis
restricting to the power-law regime.

The resulting $t/t_w$-dependent factor $\langle \tilde w^2\rangle$,
see Eq.~\eqref{eq:scaling}, can be well acquainted for by the empiric
form~\cite{EPL}
\begin{eqnarray}
&&
\langle \tilde w^2\rangle
\left(\frac{\ell(t)}{\ell(t_w)}\right) \equiv {\cal G}(x)
\qquad \mbox{with} \qquad x\equiv \Delta t/t_w
\; ,
\nonumber\\
&&
{\cal G}(x)
\sim x^{2 \alpha(T)}  A(T,T_0) \; 10^{B(T,T_0 )g(x,T,T_0 )}
\label{eq:fitting0}
\\
&&
g(x, T,T_0 ) \equiv
\tanh\left[
C(T,T_0) \log_{10} \left(\frac{x}{D(T,T_0)}\right)
\right]
\; .
\nonumber
\end{eqnarray}
Equations~\eqref{eq:scaling} and (\eqref{eq:fitting0} have the stationary limits:
\begin{eqnarray}
\langle w^2\rangle
\sim
\left\{
\begin{array}{l}
 c_0(T, T_0) \;\; \Delta t^{2 \alpha(T)} \qquad \Delta t \ll t_w
\; ,
\\
c_\infty (T, T_0) \; \Delta t^{2 \alpha(T )} \qquad \Delta t \gg t_w
\; ,
\end{array}
\right.
\label{eq:constants-c}
\end{eqnarray}
with
\begin{eqnarray}
c_0(T,T_0) &\equiv& A(T,T_0) \; 10^{-B(T,T_0)}
\; ,
\nonumber\\
c_\infty(T,T_0) &\equiv& A(T,T_0) \; 10^{B(T,T_0)}
\; .
\end{eqnarray}
These parameters control the two stationary asymptotes bounding the
aging regime.  The parameter
$2B(T,T_0)=\log_{10}[c_\infty(T,T_0)/c_0(T,T_0)]$ is then a measure of
the distance between the two asymptotes.

Figure~\ref{fig:averaged-roughness}(b) shows the scaling plot of the
averaged roughness in the growth regime. We display data for $L=5000$
at $T=0.5$, and $L=500$ at $T=0.2$. Note that the data for $L=500$ at
$T=0.5$ shown in panel (a) of Fig.~\ref{fig:averaged-roughness} would
have not scaled properly since saturation appears earlier.

\begin{figure}
\centerline{
\includegraphics[angle=0,width=.75\linewidth,clip=true]{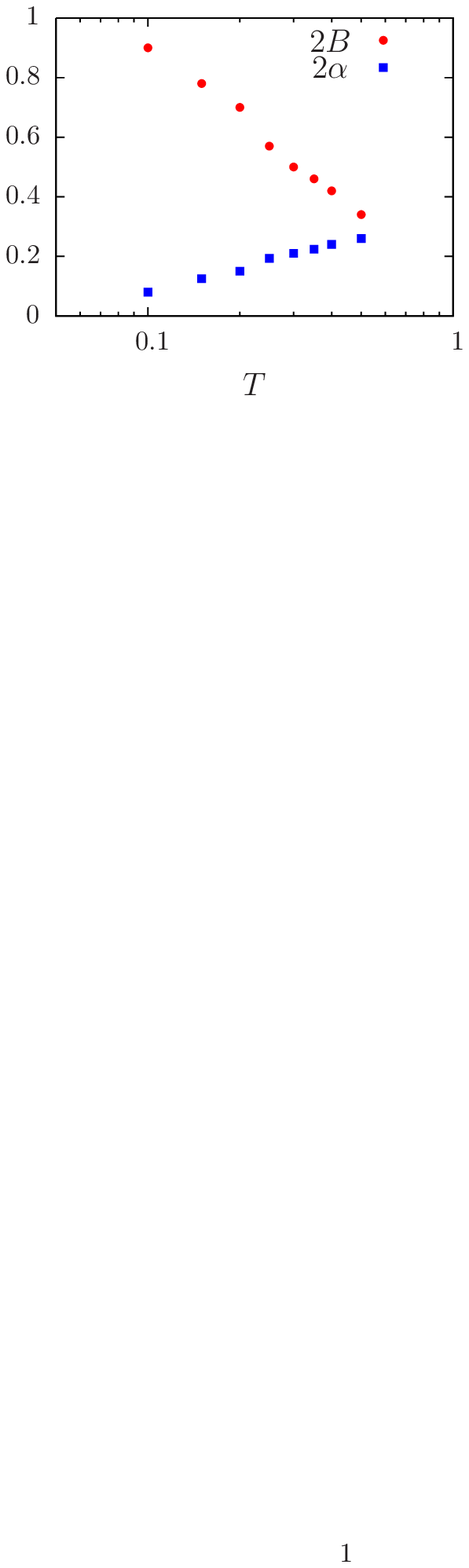}
}
\caption{(Color online)
Temperature dependence of the effective growth exponent
  $\alpha$ and the parameter $B$ measuring the distance between the
  two asymptotes $c_0$ and $c_\infty$ in the lattice model evolving
  from an infinite temperature initial condition.}
\label{fig:parameters}
\end{figure}

The temperature dependence of the (effective) growth exponent $\alpha$ as well
as the parameter $B$ for the solid-on-solid model evolving from an
infinite temperature initial condition are shown in
Fig.~\ref{fig:parameters}.  The exponent $\alpha$ approaches the EW
value $2 \alpha = 1/2$ at high temperature. In the high-temperature
limit the parameter $B(T,\infty)$ approaches zero since the two
constants $c_0(T,\infty)$ and $c_\infty(T,\infty)$ tend to the same
temperature independent value and aging disappears.

\begin{figure}
\centerline{
\includegraphics[angle=-0,width=.85\linewidth,clip=true]{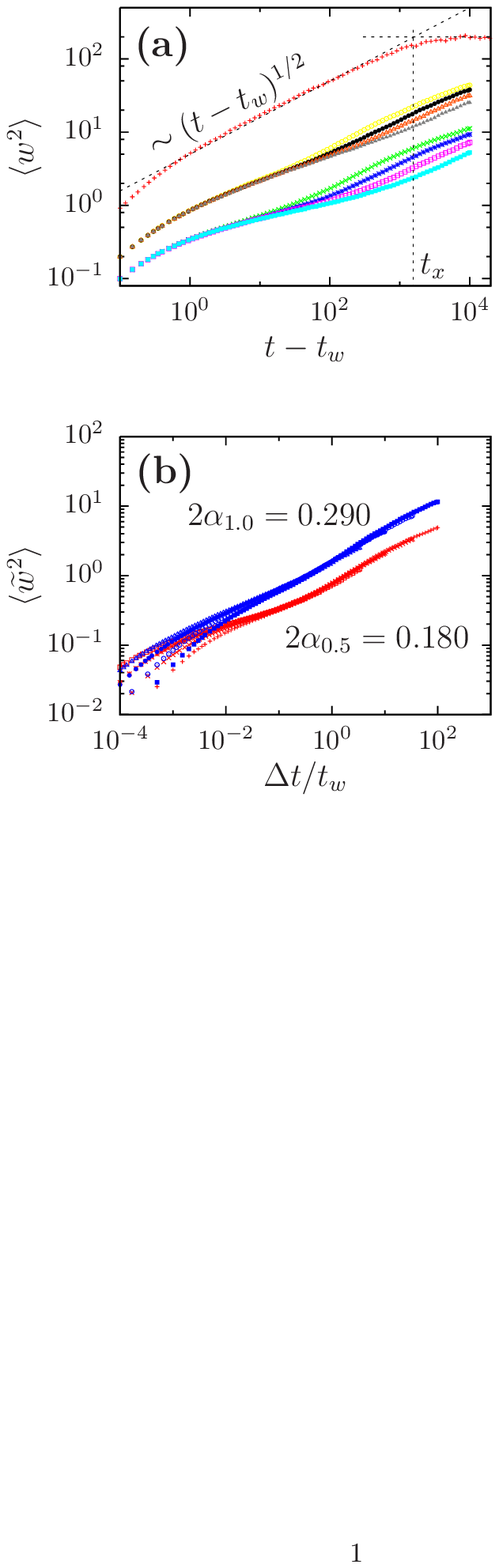}
}
\caption{(Color online) The averaged two-time roughness in the
  continuous model after a quench from high temperature $T_0=5$. (a)
  $L=256$ and different temperatures. The upper curve corresponds to
  the evolution of the roughness when the temperature is the same as
  the initial value, {\it i.e.} a stationary situation. Curves for $T=1$
  (middle curves) and $T=0.5$ (lower curves) are shonw. The waiting
  times are $t_w=100,300,1000$ and $3000$
  from left to right. (b) Scaling plot of the data presented for
  $T=1.0$ and $T=0.5$ using the $\alpha$ values given in the figure.}
\label{fig:w2_Ti5_scal}
\end{figure}

\begin{figure}
\centerline{
\includegraphics[angle=-0,width=\linewidth,clip=true]{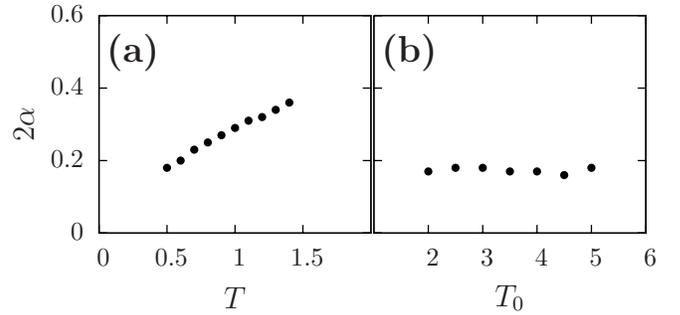}
}
\caption{Initial and working temperature dependence of the scaling
  exponent $\alpha$ in the continuous EW model in a random
  environment. (a) The initial temperature is fixed to $T_0=5$ and the
  working temperature is varied.  (b) The exponent $\alpha$ does not
  change while changing the initial temperature $T_0$ and keeping the
  working temperature fixed at $T=0.5$.}
\label{fig:alfadeT}
\end{figure}

Figure~\ref{fig:w2_Ti5_scal} shows the averaged two-time roughness for
the continuous model with a high initial temperature $T_0=5$.  The
upper curve in Fig.~\ref{fig:w2_Ti5_scal}(a) corresponds to the
stationary case where the temperature is kept constant at the initial
value. The other two cases, with $T=1$ and $T=0.5$, present aging
which is qualitatively similar to the SOS model of
Fig.~\ref{fig:averaged-roughness}. Figure~\ref{fig:w2_Ti5_scal}(b)
shows the scaling of these curves with the temperature dependent
exponents $\alpha$ given in the key.  In the continuous model the
exponent $\alpha$ increases with temperature until reaching the EW
value $2 \alpha = 1/2$ after the crossover temperature
$T_{co}$. In Fig.~\ref{fig:alfadeT}(a) we present $\alpha(T)$ for the
cases with a fixed initial temperature $T_0=5$. As can be observed,
its precise temperature dependence is not easy to determine. Since one
expects that the $\alpha$ exponent determines the slow dynamics
at a given temperature, it is interesting to check that the
value of $\alpha$ at a small temperature does not depend on the
initial temperature. This is shown in Fig.~\ref{fig:alfadeT}(b), where
we show that $\alpha(T=0.5)$ is independent of the initial
temperature.
Although we do not compute the parameter $B$ for the continuous
model, it is worth noting that in contrast with the discrete case the
two constants $c_0$ and $c_\infty$ approach a temperature dependent
asymptote $c_0\sim c_\infty\propto 2T$ that equals the high
temperature limit of the EW line, $c_0 \propto 2T$ and $c_\infty
\propto (2 - \sqrt{ 2})T + \sqrt{ 2}T_0$.~\cite{Sebastian} The exponent
$\alpha$ monotonically increases from $\alpha(T=0)\approx 0$ and crosses over
to $2 \alpha=1/2$ at some $T_{co}$ with a very weak $T$ dependence.

The effect of using as an initial configuration one
in equilibrium at a lower temperature was considered analytically
in Ref.~\onlinecite{Sebastian} for the EW line. In presence of disorder one is
often forced to use numerical simulations and the difficulty of
equilibrating a long line at low temperature arises.
However, a special case is that of the lattice model at $T_0=0$. It is
indeed well known that its ground state configuration can be exactly
calculated by transfer-matrix methods.~\cite{Kardar}

\begin{figure}
\includegraphics[width=.85\linewidth,clip=true]{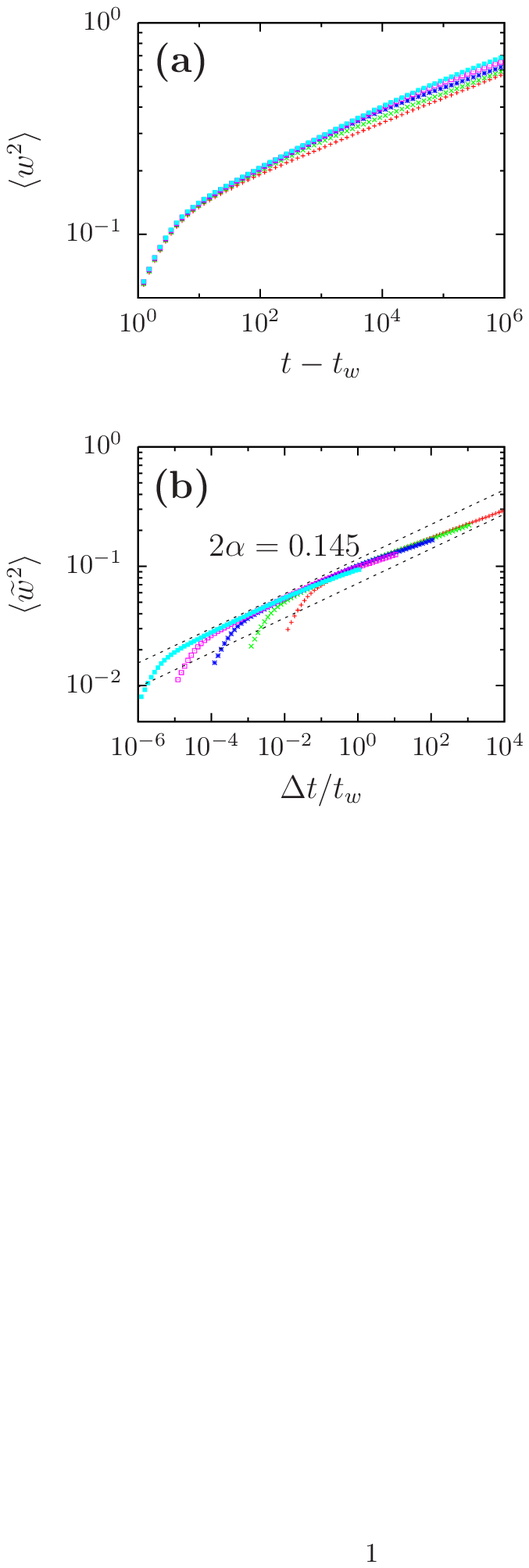}
\caption{(Color online) Panel (a): the averaged two-time roughness in
  the lattice model for an elastic line with length $L=5000$.
  Evolution at $T=0.2$ after a sudden heating from equilibrium at
  $T_0=0$. The waiting times are $t_w=10$ (plus, red), $t_w=10^2$
  (cross, green) $t_w=10^3$ (star, blue), $t_w=10^4$ (open square,
  pink), $t_w=10^5$ (filled square, cyan).  Panel (b): scaling plot of
  the data in panel (a). Note that the value of $\alpha$ (in the key)
  coincides with the one used in the case of a quench to $T$ from
  $T_0\to\infty$ (see Fig.~\ref{fig:averaged-roughness}). The dashed
  lines are the bounding asymptotes. Let us remark that, here,
  $c_\infty<c_0$.}
\label{fig:heating}
\end{figure}

In Fig.~\ref{fig:heating} we show the two-time roughness of the
discrete disordered line using as initial configurations equilibrium
ones at $T_0=0$ and a working temperature $T=0.2$. The plots
demonstrate that the trend of the curves is to reverse in the sense
that the curves go from the upper to the lower asymptote. This
behavior is similar to what was found analytically for the clean EW
line~\cite{Sebastian} and the KPZ non-linear
model.~\cite{Sebastian2} This effect is also reminiscent of what was
observed in the relaxation of the $2d$ XY model from a uniform initial
condition, corresponding to equilibrium at $T_0=0$, at finite (higher)
temperature within the spin-wave approximation.~\cite{Behose}

The effect of the initial condition can be summarized in the following
way.  At all working-temperatures that are identical to the one of the
initial condition, $T=T_0$, one finds $c_0= c_\infty$ and there is no
aging since the line is initially in equilibrium.~\cite{Sebastian}  For
$T>T_0$ one has $c_0>c_\infty$ while for $T<T_0$ one has
$c_0<c_\infty$. The values of these constants determine the
relative location of the two asymptotes of the growth regime,
$\Delta t\ll t_w$ and $\Delta t\gg t_w$.

\subsection{\label{subsec:linearresponse} The linear response}

\begin{figure}
\centerline{
\includegraphics[angle=0,width=.85\linewidth,clip=true]{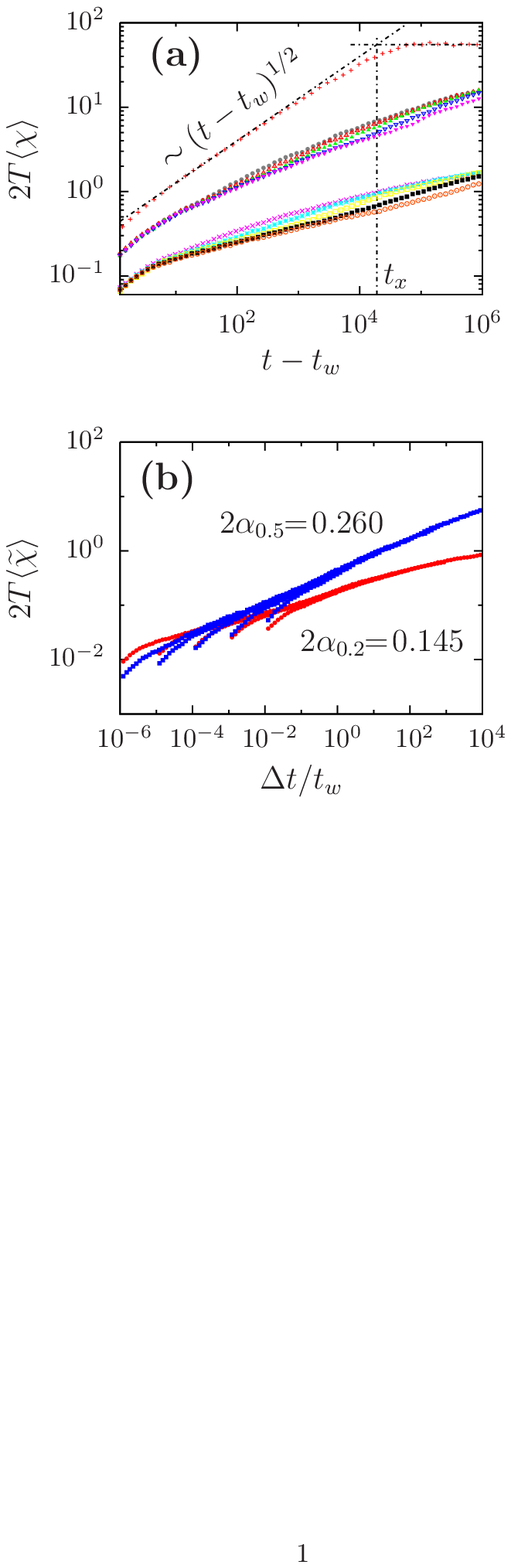}
}
\caption{(Color online) The noise and disordered averaged integrated
  linear response associated to the two-time roughness of the lattice
  elastic line.  Panel (a): the line with length $L=500$ evolving at
  $T=5, \ 0.5, \ 0.2$ after a quench from infinite temperature, $T_0
  \to\infty$.  Panel (b): scaling plot of the averaged two-time linear
  response of the line with $L=500$ and $T=0.2$, and the line with
  $L=5000$ and $T=0.5$.}
\label{fig:linear-response}
\end{figure}

We now turn to the study of the linear response.
In Fig.~\ref{fig:linear-response} we show the averaged integrated
linear response of the roughness in the lattice model. In panel (a) we
show data for several waiting-times as a function of time delay.
In panel (b) we scale the data as
\begin{equation}
\langle \chi\rangle(t,t_w) \sim \ell^{2\zeta}(t_w) \;
\langle \tilde\chi\rangle
\left(\frac{\ell(t)}{\ell(t_w)}\right)
\end{equation}
with $\ell(t)\sim  t^{1/z}$.

These results are to be confronted with the behavior of the EW
elastic line with Langevin dynamics for which the integrated linear
response is {\it stationary} and does not fluctuate.~\cite{Sebastian}
On the other hand, the
linear response obtained with the continuous model, not shown here, is
qualitatively similar to the one shown for the SOS model in
Fig.~\ref{fig:linear-response}.

\begin{figure}
\includegraphics[width=0.85\linewidth,clip=true]{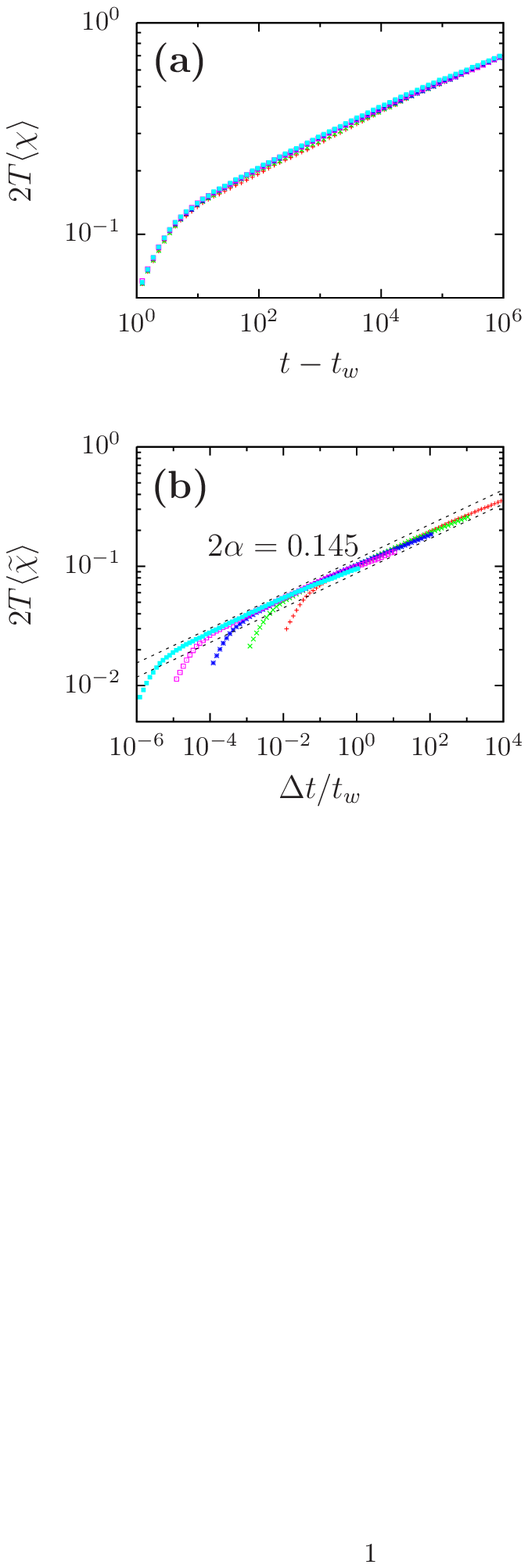}
\caption{(Color online) Panel (a): the averaged integrated linear
  response associated to the two-time roughness in the lattice model
  for an elastic line with $L=5000$.  Evolution at $T=0.2$ after a
  sudden heating from equilibrium at $T_0=0$. The waiting times and
  symbols are as in Fig.~\ref{fig:heating}.
  Panel (b): scaling plot of the data in (a). Note that the
  value of $\alpha$ (in the key) coincides with the one found in the study
  of $\langle w^2\rangle$ and the one used in the case of a quench to
  $T$ from $T_0\to\infty$ (see Figs.~\ref{fig:averaged-roughness} and
  \ref{fig:heating}). The dashed lines are the bounding
  asymptotes. Let us remark that, here, $c_\infty<c_0$.}
\label{fig:heating2}
\end{figure}

In Fig.~\ref{fig:heating2} we display data for the linear response
after the sudden heating from the ground state used in
Sect.~\ref{sec:roughness}.  The effect of this `reversed'
heating-procedure is similar to the one showed in
Sect.~\ref{sec:roughness} and in Fig.~\ref{fig:heating} for the roughness.

\subsection{\label{subsec:fdt} The Fluctuation-dissipation theorem (FDT)}

\begin{figure}
\centerline{
\includegraphics[angle=-0,width=.8\linewidth,clip=true]{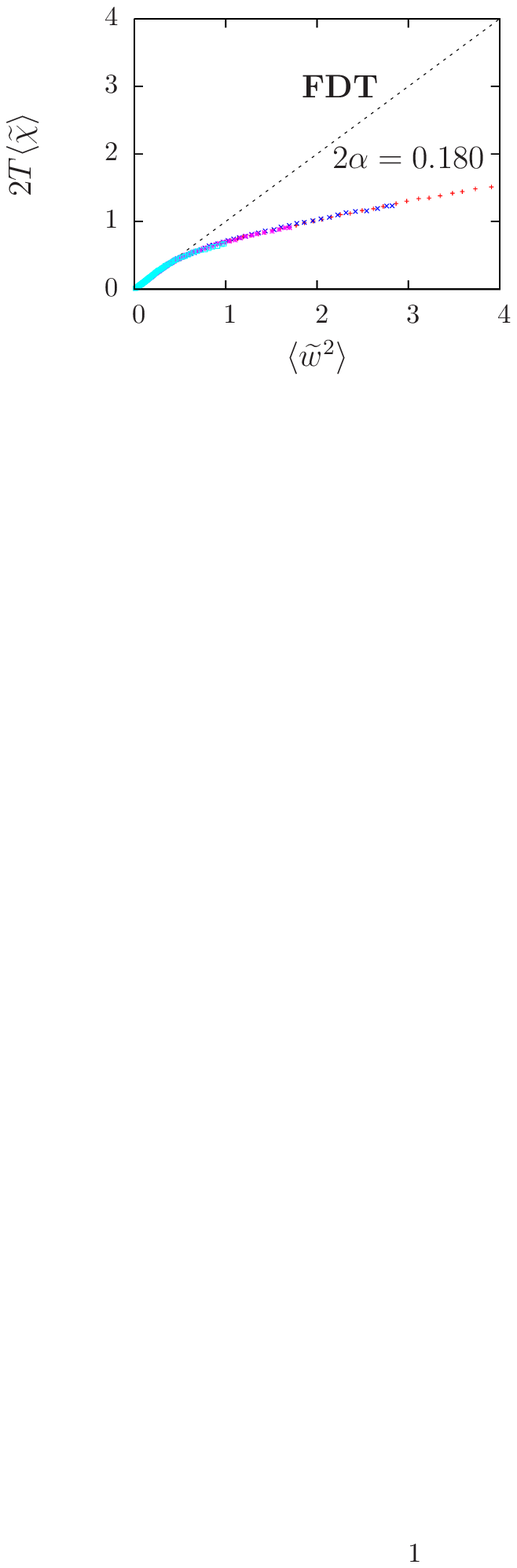}
}
\caption{(Color online) The parametric plot of the scaled linear
  response against the roughness for the continuous model with
  $T_0=5$ and $T=0.5$. The dashed line indicates the FDT limit.}
\label{fig:w2yX_Ti5_Tf0.5_F01_FDT}
\end{figure}

The modification of the FDT linking the displacement to its associated
response in the lattice model after a quench from infinite temperature
was studied by A. Barrat~\cite{Barrat} and
Yoshino.~\cite{Yoshino,Yoshino-unp} We here focus on the behavior of
the roughness and its linear response. We show data for the disordered
EW line after a similar quench.  In
Fig.~\ref{fig:w2yX_Ti5_Tf0.5_F01_FDT} we show the plot $\langle \tilde
\chi\rangle$ against $\langle \tilde w^2\rangle$ at fixed $t_w$ and
using $\Delta t$ as a parameter going from $\Delta t=0$ to $\Delta
t\to\infty$ in the continuous disordered EW line for the case $T_0=5$
and $T=0.5$. As it can be observed, it displays two slopes,
allowing for the definition of an effective temperature~\cite{Cukupe}.
A similar behaviour was reported for the displacement and associated
linear response in the lattice model,~\cite{Yoshino} the
clean EW line,~\cite{Sebastian} and the vortex glass model.~\cite{us}

\begin{figure}
\centerline{
\includegraphics[angle=-0,width=\linewidth,clip=true]{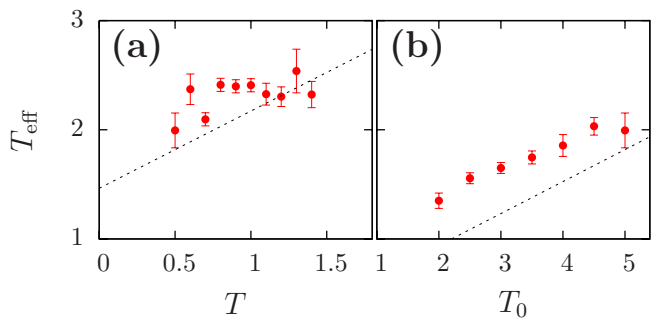}
}
\caption{(Color online) Temperature dependence of the effective
  temperature, signaling the violation of the FDT, comparing the clean
  EW system --dashed lines-- with the disordered continuous case. (a)
  Dependence on the working temperature for a fixed initial
  temperature $T_0=5$. (b) Dependence on the initial temperature $T_0$
  while the working temperature is kept fixed at $T=0.5$.}
\label{fig:TeffdeT}
\end{figure}

The temperature dependence of the effective temperature is shown in
Fig.~\ref{fig:TeffdeT} for the disorder and the clean case. The
dependence of $T_{\rm eff}$ on the working temperature while keeping $T_0$
fixed is shown in Fig.~\ref{fig:TeffdeT}(a), but it is difficult to
asses whether the effective temperature is constant or slowly grows
with $T$. Without disorder the EW solution gives a linear dependence
$T_{\rm eff}=T/\sqrt{2}+T_0(1+1/\sqrt{2})$,~\cite{Sebastian} shown with
dashed lines in the figure. When the working temperature is fixed at
$T=0.5$ and the initial temperature is changed, $T_{\rm eff}$ grows
linearly with $T_0$, as shown in Fig.~\ref{fig:TeffdeT}(b) and
in the clean limit.

\begin{figure}
\centerline{
\includegraphics[angle=-0,width=.8\linewidth,clip=true]{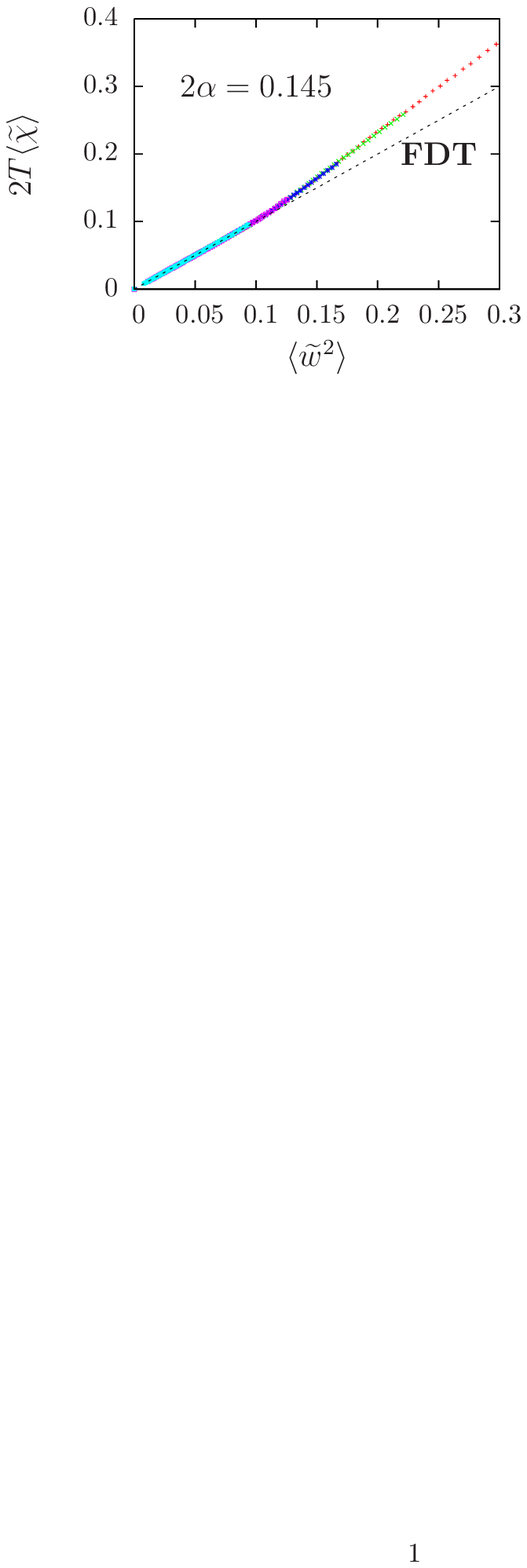}
}
\caption{(Color online) The parametric plot $2T\langle \widetilde
  \chi\rangle$ against $\langle \widetilde w^2\rangle$ after a heating
  procedure from the ground state at $T_0=0$ to $T=0.2$ in the SOS
  model.  The waiting times are $10$ MCs (plus, red), $10^2$ MCs
  (cross, green), $10^3$ MCs (star, blue), $10^4$ MCs (open square,
  pink) and $10^5$ MCs (filled square, cyan).  Note that $T_{\rm eff}<T$.}
\label{fig:TeffdeT-heating}
\end{figure}

We also studied the effect of a heating procedure on the violations of
the FDT using the lattice model. We show in
Fig.~\ref{fig:TeffdeT-heating} the parametric plot $\langle \tilde
\chi\rangle$ against $\langle \tilde w^2\rangle$ using as initial
condition the ground state at $T=0$. As already found in the clean EW
line we find that $T_{\rm eff}<T$. $T_{\rm eff}$ thus reflects, consistently,
a `memory' of the initial configuration.

\section{\label{sec:discusion} Crossover-induced
geometrical and dynamical effects}

We discuss here how our numerical results can be explained by developing
simple scaling arguments based on the existence of
a single dynamic crossover in the growing correlation length at
a static temperature dependent length $L_T$, from a
thermally dominated regime, to a disorder dominated regime
with algebraically growing barriers as function of the
length-scale. The main additional assumption
on that we make is
that at large enough length-scales $\ell \gg L_T$, the
geometry and typical barriers controlling the dynamics are
indistinguishable from those at zero temperature.  With these
hypothesis, which are qualitatively supported by our numerical results,
we can predict: (i) the temperature dependence of
$L_T$; (ii) the temperature dependence of the effective
exponents characterizing each regime of growth; (iii)
the temperature dependence of the crossover time between these regimes;
(iv) a parameter quantifying the importance of crossover-induced
finite-size/time effects in simulations. In order to
make the discussion in this Section self-contained we start by
repeating some arguments that are well-known in the literature and
then we present the scenario that, we propose, explains the numerical
observations.

If the temperature is high enough to renormalize the microscopic
pinning parameters,~\cite{Nattermann} thermal fluctuations dominate the line wandering
at short length-scales. The geometry of equilibrated small scales,
$\ell \ll L_T$, is thus described by the thermal roughness exponent
$\zeta_T=1/2$ of the clean Edwards-Wilkinson (EW) equation, such that
$\langle w^2\rangle \sim (T/c) \ell$ for $ \ell/L_T \ll 1$ or
$S(q)\sim (T/c) q^{-(1+2\zeta_T)}$ for $qL_T \gg 1$, or $q \gg q_T
\sim 1/L_T$.  The typical time needed to equilibrate a length $\ell$
in this regime is therefore expected to be $t(\ell) \sim \ell^{z}$,
with the EW dynamical exponent $z=2$.

At length-scales $\ell \gg L_T$, disorder dominates and the geometry
of the equilibrated large length scales, $\ell \gg L_T$, is the same as
in the ground-state, $S(q) \sim q^{-(1+2\zeta_D)}$ for $q L_T \ll 1$,
with $\zeta_D =2/3 > \zeta_T$. The typical time to equilibrate a
length $\ell \gg L_T$ is controlled by size-dependent {\it energy}
barriers $U(\ell)$, such that $t(\ell) \sim e^{U(\ell)/T}$, with
$U(\ell) \sim \ell^{\psi} \gg T$.  By hypothesis, we assume that both
the form of $S(q)$ and the barriers $U(\ell)$ are independent of $T$
in this regime, in agreement with what is seen in
Fig.~\ref{fig:pru_1cTf0.5} for the structure factor with $q \ll q_T$.

In order to capture the crossover effects we interpolate the
geometric and dynamic behavior of the regimes $\ell \gg L_T$ and
$\ell \ll L_T$ described above.  The static structure factor can be
written as
\begin{equation}
S(q,T) \sim q^{-(1+2\zeta_D)} g(q L_T, T)
\end{equation}
with $g(x, T)\sim T x^{2(\zeta_D - \zeta_T)}$ for $x \gg 1$ and
$g(x,T)\sim \mbox{const.}$ for $x \ll 1$.  This interpolating form,
which includes the temperature, is found to fit well the data, and to
satisfy our assumption $S(q, T) \approx S(q, T=0)$ for $q L_T \ll 1$.
We continuously match the two regimes by requiring  $q_T^{-(1+2 \zeta_D)} = T
q_T^{-(1+2 \zeta_T)}$ and, therefore,
\begin{equation}
L_T \sim T^{1/\phi} = T^3,
\end{equation}
with $\phi = 2(\zeta_D-\zeta_T) = 1/3$, which is the temperature
dependence of the crossover length we observe in
Fig. \ref{fig:pru_qc}. This argument was also discussed in
Sect. \ref{sec:crossoverlength} on numerical grounds, where references
were also given.

Let us now discuss the dynamical behavior induced by $L_T$.  At the
crossover we can roughly set $U(L_T) \sim T$, since barriers start to dominate
the dynamics only above $L_T$.
For $\ell\gg L_T$ we expect $U(\ell) \sim \ell^{\psi} \gg T$.  To
interpolate the large scale behavior down to $L_T$ we use~\cite{Nattermann}
\begin{equation}
U(\ell)\sim T (\ell/L_T)^\psi.
\end{equation}
Since by hypothesis $U(\ell,T) \approx U(\ell, T=0)$ for $\ell\gg
L_T$, we obtain $\psi = \phi ={2(\zeta_D-\zeta_T)} \equiv \theta =
1/3$, with $\theta$ the energy exponent. The time to equilibrate a
length $\ell$ can hence be written as
\begin{equation}
t(\ell) \sim e^{(\ell/L_T)^\psi} \ell^z.
\label{eq:tvsl}
\end{equation}
This expression continuously matches the barrier-dominated regime,
$t \sim \exp[(\ell/L_T)^\psi] =
\exp[\ell^\psi/T]$ for $\ell \gg L_T$, with the power-law growth
expected in the thermal regime, $t \sim \ell^z$ for $\ell \ll L_T$.
From Eq. \eqref{eq:tvsl} and using $L_T \sim T^{1/\phi}$ one obtains a
characteristic time scale,
\begin{equation}
t_{\times} \sim L_T^z \sim T^{z/\psi} = T^6,
\end{equation}
separating thermally from disorder dominated dynamics. This strong
temperature dependence explains why the crossover to the slow
logarithmic growth is particularly difficult to observe in numerical
simulations. On the one hand, at high temperature, the line size must
be large $L > L_T \sim T^3$ in order to see the crossover to the
long-time logarithmic regime. On the other hand, at low temperature,
the lines could in principle cross over to the logarithmic regime even
for small system sizes but, because of the very slow Arrhenius
activated dynamics, exceedingly long running times would be needed to
resolve the precise time dependence. Most importantly, the data from
any numerical simulation displaying the two regimes of growth will be
characterized by temperature-dependent effective exponents due to
crossover-induced finite-size/time effects, as we explain below.

Indeed, Eq.~\eqref{eq:tvsl} allows us to explain the temperature
dependence in the effective exponents of the power-law and logarithmic
growth regimes.  The influence of the crossover in the effective
dynamical exponent $\tilde z$ of the power-law regime relevant to short
length-scales, $\ell \ll L_T$, follows from equating Eq.~\eqref{eq:tvsl}
to $t(\ell) \sim \ell^{\tilde z}$, which yields
\begin{equation}
{\tilde z} - z \approx  \frac{(\ell^\psi/\ln \ell)}{L_T^{\psi}} \approx \frac{(\tilde z \; t^{\psi/\tilde z}/\ln t)}{L_T^{\psi}}.
\end{equation}
To avoid solving the resulting self-consistent equation,
we can  use $\tilde z \approx z$ in the third member to obtain a first order correction to the dynamical exponent,
\begin{equation}
{\tilde z}-z \approx \frac{(z\; t^{\psi/z}/\ln t)}{L_T^{\psi}}.
\end{equation}
Since by construction the power-law fit of the data is done in a
relatively short time-scale such that $t^{\psi/z}/\ln t$ is
almost constant (otherwise $\tilde z$ would not be well defined),
and by using $L_T\sim T^3$ and $\psi= 1/3$, we get
\begin{equation}
\tilde z - 2 \simeq  T^{-1}.
\end{equation}
This rough estimate is
in good agreement with the numerically determined
exponent $\alpha$ shown in Figs.~\ref{fig:parameters} and~\ref{fig:alfadeT}.

It is worth noting here that the Cardy-Ostlund random Sine-Gordon model~\cite{Greg} in
$d=2$, disordered ferromagnets,~\cite{Pleimling-z}
and Ising and vector spin glasses~\cite{Helmut}
were shown to display a dynamic exponent $z \sim 2+ a/T$, with $a$
of the order of unity and depending on details of the
considered model. Our arguments and numerical study show that
the numerically measured dynamic exponent
could simply be an effective value that stems from the
interpolation of two growth regimes that cannot be
properly resolved numerically.
This seems to be particularly important for systems with a positive
energy exponent $\theta$, since in general we expect
$\psi \sim \theta$, and
therefore a crossover from a thermally dominated short-time regime
to
a long-time dynamics dominated by diverging energetic barriers.

The temperature dependence of the effective exponent in the regime of
logarithmic growth, $\ell \gg L_T$, can be explained with a similar argument.
By equating Eq.~\eqref{eq:tvsl} to
$t(\ell) \sim \exp [\ell^{\tilde \psi}/T]$ we find,
\begin{equation}
\tilde \psi - \psi
\approx
\frac{\ln [T/L_T^\psi +  zT \ln \ell/\ell^\psi]  }{ \ln \ell}.
\end{equation}
We use  now  $L_T\sim T^3$ and
the time-dependence $\ell \sim (T\ln t)^{1/\tilde \psi}$ in the
second member  to get a self-consistent equation
for $\tilde \psi$, namely
\begin{equation}
\tilde \psi - \psi  \sim
 \frac{\tilde \psi}{\ln (T \ln t)} \ln \left[ 1 +
  \frac{Tz}{\tilde \psi}
   \frac{\ln (T \ln t)}{(T \ln t)^{\psi/\tilde \psi}} \right].
\end{equation}
Fixing the time scale and setting $\tilde \psi \approx \psi$ in the second member we obtain a first order approximation for $\tilde \psi - \psi$,
\begin{equation}
\tilde \psi - \psi  \sim
 \frac{\psi}{\ln (T \ln t)} \ln \left[ 1 +  \frac{z}{\psi}
  \frac{\ln (T \ln t)}{\ln t} \right].
\end{equation}
At high temperatures
(or large times) such that $T \ln t \gg 1$ this expression simply yields
$\tilde \psi \sim \psi$.
In the opposite low $T$ limit for sufficiently large length scales such
that the restriction $\ell \gg L_T\sim T^3$ is still satisfied one
finds $\tilde \psi > \psi$.  Indeed, $\tilde \psi$
is a decreasing function of $T$.
These results are consistent with the result shown in Fig.~\ref{fig:theta-fit}, where we observe that
$\tilde \psi > \psi$ at low temperatures, and
$\tilde \psi \to \psi=1/3$ increasing the temperature.

The effective exponent analysis made above reveals that crossover
induced effects are controlled by the order of magnitude of
$\ell^{\psi}/\ln \ell$. Since $\ell$ in a simulation is always a
fraction of the system size $L$, a temperature dependence of the
effective exponents can be expected for systems such that $L^{1/3}/\ln
L \sim O(1)$.  For instance, in order to properly resolve the dynamic
exponent $z=2$ at short length scales we need $\ell \ll L_T$ for at
least a few orders of magnitude range of $\ell$, {\it i.e.}  a large
$L_T$. On the other hand, to properly resolve the exponent $\psi$ we
need $\ell^{1/3}/\ln \ell \gg 1$ for $L > \ell > L_T$ for at least a
few orders of magnitude of $\ell$. Typically, the largest systems
analyzed numerically in the literature have $L \sim O(10^3)$, with
$L^{1/3}/\ln L \sim O(1)$ and do not satisfy the latter constraint.
This clearly prompts for a careful interpretation of the anomalous
temperature dependence of power-law or logarithmic growth
exponents obtained numerically so far, as we discuss below.

In  Ref.~\onlinecite{Kolton}  the  relaxation   of  an  elastic  line  in  a
disordered potential was analyzed  for the particular case $t_w=0$ and
$T=0.5$. The crossover to  the logarithmic growth was observed, though
with an  effective exponent $\tilde  \psi \approx 0.49 >  \theta$.  In
Ref.~\onlinecite{Monthus}  scaling  arguments for  the  free  energy of  droplet
excitations  were given  favoring  the value  $\psi=1/2$, against  the
commonly  accepted  value  $\psi  = \theta  =  1/3$,~\cite{Kardar}  in
apparent  agreement  with the  results  of  Ref.~\onlinecite{Kolton}. In  the
present work we find $\tilde  \psi \sim 0.51$ for $t_w=0$ and $T=0.5$,
consistent with  the result of Ref.~\onlinecite{Kolton}, but  also a decrease
towards a  value close to $\psi=1/3$ with  temperature. This behavior
is opposite to what is proposed in Ref.~\onlinecite{Monthus}, where entropic
effects leading to the value $\psi=1/2$ are expected to grow with $T$,
from  the $T=0$  value $\psi=\theta=1/3$,  to  the entropically-driven
value $\psi=1/2$. The scaling arguments given above thus show that the
anomalous  temperature  dependence  of  $\tilde  \psi$  can  still  be
interpreted without dropping the $\psi=\theta=1/3$ identification,
which was proposed time ago.~\cite{Kardar}

In numerical simulations of the steady-state slow driven motion of
an elastic line ~\cite{Kolton-creep} the effective creep exponent $\mu$ was
found to be temperature dependent. According
to the assumption $U(\ell)\sim \ell^\theta$ at large length-scales, simple
scaling arguments lead to $\mu = \theta/(2-\zeta_D)=1/4$
(see, for instance Ref.~\onlinecite{giamarchi-kolton-rosso}). Taking into account
that $L^{1/3}/\ln L \sim O(1)$ in those simulations we can
expect, following an equivalent line of reasoning as above, a
 temperature dependent
effective creep exponent $\tilde \mu = \tilde \psi/(2-\zeta_D) > \mu$,
such that $\tilde \mu \to \mu$ when $T$ grows. This behavior is in good
qualitative agreement with the results in Ref.~\onlinecite{Kolton-creep}.
Therefore, the anomalous temperature dependence observed in Ref.~\onlinecite{Kolton-creep}
can not be used as evidence against obtaining the creep exponent entirely as
a function of static exponents, although the large scale geometry
in the creep regime is found to be described
by depinning exponents.~\cite{Kolton-creep,Kolton-vmcreep}

\section{\label{sec:conclusions} Conclusions}

Elastic manifolds are objects that appear in a large variety of
problems with glassy features. It was shown in the past that these
systems present aspects of glassy dynamics combined with diffusion
properties.~\cite{Cule,Yoshino,Barrat,Yoshino-unp,us,EPL,Sebastian,Pleimling,Sebastian2}

After a series of analytic and numerical studies of continuous and
discrete models with and without quenched disorder we can
summarize the main features of the out of equilibrium relaxation
of individual and interacting  directed one-dimensional objects.

All these systems age below a characteristic temperature $T_{co}$. In
the mean-field limit of an infinite number of transverse dimensions,
for a line with infinite length, a dynamic phase
transition arise at $T_{co}$. Below $T_{co}$ the lines age without
diffusion.~\cite{Cule}

The dynamics of finite length lines moving in finite dimensional
spaces cross over from diffusive-aging growth to a regime in which the
roughness saturates.  This phenomenon can be described with a
generalization of the Family-Vicsek scaling.~\cite{us,Sebastian} The
qualitative features of the two-time freely relaxing observables are
generic and do not depend on the presence of quenched randomness but
the details such as the exponents and growing length do. We performed
a careful analysis of the time-dependent growing length and we found
a crossover from an effective power-law with temperature dependent
exponent to the expected asymptotic logarithmic growth for
sufficiently long strings. Our numerical data indicates a temperature
dependent effective barrier exponent $\tilde \psi$, which is higher
than the energy exponent $\theta=1/3$ at low temperatures, but tends
to it at high temperatures, provided the system is large enough to
display the asymptotic behavior.  As we discussed via scaling
arguments this behavior can be recast in terms of crossover
induced effects, and it is not inconsistent with the usual
assumption $\psi = \theta$ for this system.

We also found a single crossover length $L_T \sim T^3$
separating a thermally dominated regime with $\zeta_T=1/2$
and a disorder dominated regime with the $T=0$ roughness exponent
$\zeta_D = 2/3$. We showed that this is consistent with the
assumption that large scale properties of the system are
indistinguishable from those at $T=0$. Translating this crossover
into the dynamics we could also describe the temperature dependence
of the effective dynamical exponents in the power-law growth regime,
and in the logarithmic growth regime.
In this respect it would be important to study the corresponding
crossover in $(1+2)$ dimensions, relevant for vortices in
superconductors. This is crucial for high-Tc
superconductors where thermal fluctuations are known to strongly
renormalize the disorder and produce Larkin lengths growing exponentially
fast with temperature.~\cite{rusos} As the Larkin length marks
the onset of metastability and the crossover to a barrier dominated
(random-manifold) regime, as it does $L_T$ in our case,
we can expect the phenomenology and crossover induced effects we describe here
to be experimentally relevant, specially at the onset of irreversibility,
near the transition to the vortex liquid phase,~\cite{rusos,us} where
the Larkin length can become comparable with the sample size.~\cite{moira}

The importance of measuring linear responses was demonstrated by the
fact that all known coarsening systems (above their lower critical
dimension) have an asymptotically vanishing linear response in the
aging regime, in contrast to solvable mean-field glassy models and
numerical simulations of finite dimensional glasses that yield a
finite integrated linear response in the same two-times regime (see {\it
e.g.} Refs.~\onlinecite{Leto} and \onlinecite{Corberi}). This fact appears as a
concrete difference between the relaxation dynamics of coarsening
and glassy systems.

The relaxation of the integrated thermal averaged linear
response of elastic manifolds strongly depends on the presence or
absence of quenched disorder.  Mean-field elastic manifold models in
quenched random environments have a fully aging linear response. The
clean EW line has a {\it stationary} linear response while the dirty
continuous or lattice models have integrated linear responses with
aging and diffusion combined.

The effective temperature~\cite{Cukupe} defined from the comparison of
induced and spontaneous averaged fluctuations is {\it finite} in all
the cases considered as long as the diffusive factors in the
correlations and linear responses are divided away. We also showed
that the effective temperature is higher than the bath temperature for
cooling procedures and, inversely, it is lower than the bath
temperature for heating procedures. This is similar to what has been
previously found in the $2d$ XY model~\cite{Behose} and the
Edwards-Wilkinson elastic line~\cite{Sebastian} and gives support to
the notion of effectvie temperature as measured from deviations from
the fluctuation-dissipation relation.

The study of dynamic fluctuations in these systems has not been fully
developed yet. In Ref.~\onlinecite{EPL} we analyzed the fluctuations of the
two-time roughness of the lattice disordered model during growth.  The
equilibrium~\cite{Racz} and out of equilibrium~\cite{Sebastian}
roughness fluctuations in the EW line show a similar pattern.  The
probability distribution functions in all these models satisfy a
scaling law and the scaling function follows the same trend as
a function of all its variables.  The question then arises as to whether
the fluctuations of the linear responses of clean and disordered
elastic lines are similar or different.  We shall discuss this
problem in a separate publication. The relation with the
theory of fluctuations based on time-reparametrization
invariance~\cite{Chamon-Cugliandolo} will also be discussed elsewhere.

\vspace{0.5cm}
\noindent
\underline{Acknowledgments.}
We thank D. Dom\'{\i}nguez, T. Giamarchi,
G. Schehr and H. Yoshino for very useful discussions.
LFC thanks the Universidad
Nacional de Mar del Plata, Argentina; SB and JLI thank the
Laboratoire de Physique Th\'eorique et Hautes Energies, Universit\'e
Pierre et Marie Curie Paris VI, for hospitality during the preparation
of this work. SB also thanks the University of Gen\`eve and the DPMC, where part of this
work was initiated and the Swiss NSF under MaNEP and Division II for support. The authors
acknowledge financial support from ANPCyT-PICT20075, CONICET PIP5648
(JLI), and SECyT-ECOS P. A01E01/A08E03, and PICS 3172 (LFC).
LFC is a member of Institut Universitaire de
France.


\begin{thebibliography}{99}


\bibitem{Barabasi-Stanley}
A-L Barab\'asi and H. E. Stanley, 1995 {\it Fractal concepts in surface growth} (Cambridge: Cambridge University Press);
T. Halpin-Healey and Y-C Zhang, Phys. Rep. {\bf 254}, 215 (1995).

\bibitem{Alan}
A. J. Bray, Adv. Phys. {\bf 43}, 357 (1994).

\bibitem{cracks}
A. Hansen, E. L.  Hinrichsen and S. Roux, Phys. Rev. Lett. {\bf 66}, 2476 (1991);
E. Bouchaud, J. Phys. Chem. {\bf 9}, 4319 (1997);
M. Alava, P. K. V. V. Nukalaz and S. Zapperi, Adv. Phys. {\bf 55}, 349 (2006).

\bibitem{polymers}
L. C. E. Struick, {\it Physical aging in amorphous polymers and other materials} (Elsevier, Amsterdam, 1978).

\bibitem{rusos}
G. Blatter, M. V. Feigel'man, V. B. Geshkenbein, A. I. Larkin, and V. M. Vinokur, Rev. Mod. Phys. {\bf 66}, 1125 (1994);
T. Nattermann and S. Scheidl, Adv. Phys. {\bf 49}, 607 (2000);
T. Giamarchi and P. Le Doussal, in {\it Spin glasses and random fields}, A. P. Young ed. (World Scientific, Singapore, 1997).

\bibitem{us}
S. Bustingorry, L. F. Cugliandolo and D. Dom\'{\i}nguez, Phys. Rev. Lett. {\bf 96}, 027001 (2006);
S. Bustingorry, L. F. Cugliandolo, and D. Dom\'{\i}nguez, Phys. Rev.  B {\bf 75}, 024506 (2007).

\bibitem{Cukupe}
L. F. Cugliandolo, J. Kurchan, and L. Peliti, Phys. Rev. E {\bf 55}, 3898 (1997).

\bibitem{Leto}
L. F. Cugliandolo, in {\it Slow Relaxations and Nonequilibrium Dynamics in Condensed Matter} ({\it Ecole de Physique Les Houches vol 77}), ed J-L Barrat {\it et al.} (Berlin: Springer, 2003). Also available as [cond-mat/0210312].

\bibitem{high-Tc}
F. Portier, G. Kriza, B. Sas, L. F. Kiss, I. Pethes, K. Vad, B. Keszei, and F. I. B. Williams, Phys. Rev. B {\bf 66}, 140511(R) (2002);
R. Exartier and L. F. Cugliandolo, Phys. Rev. B {\bf 66}, 012517 (2002).

\bibitem{low-Tc}
X. Du, G. Li, E. Y. Andrei, M. Greenblatt, and P. Shuk, Nature Phys. {\bf  3}, 111 (2007).

\bibitem{Cule}
L. F. Cugliandolo and P. Le Doussal, Phys. Rev. E {\bf 53}, 1525 (1996);
L. F. Cugliandolo, J. Kurchan, and P. Le Doussal, Phys. Rev. Lett. {\bf 76}, 2390 (1996);
Z. Konkoli, J. Hertz, and S. Franz, Phys. Rev. E {\bf 64}, 051910 (2001);
Z. Konkoli and J. Hertz, Phys. Rev. E {\bf 67}, 051915 (2003);
Y. Y. Goldschmidt, Phys. Rev. E {\bf 74}, 021804 (2006).


\bibitem{Barrat}
A. Barrat, Phys. Rev. E {\bf 55}, 5651 (1997).

\bibitem{Yoshino}
H. Yoshino, J. Phys. A  {\bf 29}, 1421 (1996); Phys. Rev. Lett. {\bf 81}, 1493 (1998).

\bibitem{Yoshino-unp}
H. Yoshino, unpublished.

\bibitem{EPL}
S. Bustingorry, J. L. Iguain, C. Chamon, L. F. Cugliandolo, and D. Dom\'{\i}nguez, Europhys. Lett. {\bf 76}, 856 (2006).

\bibitem{Favi}
F. Family and T. Vicsek, J. Phys. A {\bf 18}, L75 (1985).

\bibitem{EW}
S. F. Edwards and D. R. Wilkinson, Proc. R. Soc. London Ser. A {\bf 381}, 17 (1982).

\bibitem{Sebastian}
S. Bustingorry, L. F. Cugliandolo, and J. L. Iguain, J. Stat. Mech. (2007) P09008.

\bibitem{Pleimling}
A. R\"othlein, F. Baumann, and M. Pleimling, Phys. Rev. E {\bf 74}, 061604 (2006); Phys. Rev. E {\bf 76}, 019901(E) (2007).

\bibitem{Sebastian2}
S. Bustingorry, J. Stat. Mech. (2007) P10002.

\bibitem{Spaniards}
See also J. J. Ramasco, J. M. L\'opez and M. A. Rodr\'{\i}guez, Europhys. Lett. {\bf 76}, 554 (2006).

\bibitem{KPZ}
M. Kardar, G. Parisi, and Y-C Zhang, Phys. Rev. Lett. {\bf 56}, 889 (1986).

\bibitem{claudio}
A. Rahmani, C. Castelnovo, J. Schmit, and C. Chamon, J. Stat. Mech. (2007) P09022.

\bibitem{Davide}
D. Loi, S. Mossa, and L. F. Cugliandolo, Phys. Rev. E {\bf 77}, 051111 (2008).

\bibitem{rosso_depinning_simulation}
A. Rosso and W. Krauth, Phys. Rev. E {\bf 65}, 025101(R) (2002).

\bibitem{Nattermann}
T. Nattermann, Y. Shapir, and I. Vilfan, Phys. Rev. B {\bf 42}, 8577 (1990).

\bibitem{Kolton-us}
A. B. Kolton, R. Exartier, L. F. Cugliandolo, D. Dom\'{\i}nguez and N. Gronbech-Jensen, Phys. Rev. Lett.
{\bf 89}, 227001 (2002).

\bibitem{Kardar}
M. Kardar, Phys. Rev. Lett. {\bf 55}, 2923 (1985);
B. Drossel and M. Kardar, Phys. Rev. E {\bf 52}, 4841 (1995).

\bibitem{drossel} B. Drossel and M. Kardar, Phys. Rev. E {\bf 52}, 4841 (1995).

\bibitem{Kolton}
A. B. Kolton, A. Rosso, and T. Giamarchi, Phys. Rev. Lett. {\bf 95}, 180604 (2005).

\bibitem{Yosh07}
T. Nogawa, K. Nemoto, and H. Yoshino, Phys. Rev. B  {\bf 77}, 064204 (2008).

\bibitem{Monthus}
C. Monthus and T. Garel, J. Phys. A {\bf 41}, 115002 (2008).

\bibitem{Behose}
L. Berthier, P. W. C. Holdsworth, and M. Sellitto, J. Phys. A {\bf 34}, 1805 (2001).

\bibitem{Greg}
G. Schehr and P. Le Doussal, Phys. Rev. E {\bf 68}, 046101 (2003);
G. Schehr and P. Le Doussal, Phys. Rev. Lett. {\bf 93}, 217201 (2004);
G. Schehr and P. Le Doussal, Europhys. Lett. {\bf 71}, 290 (2005);
G. Schehr and H. Rieger, Phys. Rev. B {\bf 71}, 184202 (2005).

\bibitem{Pleimling-z}
M. Henkel and M. Pleimling, Europhys. Lett. {\bf 76}, 561 (2006);
M. Henkel and M. Pleimling, Phys. Rev. B {\bf 78}, 224419 (2008).

\bibitem{Helmut}
H. G. Katzgraber and I. A. Campbell, Phys. Rev. B {\bf 72}, 014462 (2005).

\bibitem{Kolton-creep}
A. B. Kolton, A. Rosso, and T. Giamarchi, Phys. Rev. Lett. {\bf 94}, 047002 (2005).

\bibitem{Kolton-vmcreep}
A. B. Kolton, A. Rosso, T. Giamarchi, and W. Krauth, Phys. Rev. Lett. {\bf 97} 057001 (2006).

\bibitem{Corberi}
F. Corberi, E. Lippiello and M. Zannetti, J. Stat. Mech. (2004) P12007.

\bibitem{Racz}
Z. R\'acz, SPIE Proceedings {\bf 5112}, 248 (2003).

\bibitem{Chamon-Cugliandolo}
C. Chamon and L. F. Cugliandolo, J. Stat. Mech. (2007) P07022.

\bibitem{fisher-huse}
D.S. Fisher and D.A. Huse, Phys. Rev. B {\bf 38}, 386 (1988);
Phys. Rev B {\bf 38}, 373 (1988); Phys. Rev. B {\bf 43}, 10728 (1991).

\bibitem{giamarchi-kolton-rosso}
T. Giamarchi, A. B. Kolton, A. Rosso, Lecture Notes in Physics {\bf 688}, 91 (2006).

\bibitem{moira} M. I. Dolz, H. Pastoriza (private communication).

\end{thebibliography}
\end{document}